\documentclass[12pt]{article}
\pdfoutput=1
\usepackage{xcolor}
\usepackage{cite}
\usepackage{hyperref}
\usepackage{subcaption}
\usepackage[utf8]{inputenc}
\usepackage{graphicx}
\usepackage{kbordermatrix}

\usepackage{setspace}
\usepackage{amsmath, amssymb, amsthm, float, graphicx}
\numberwithin{equation}{section}

\hypersetup{colorlinks=false, linkcolor=blue, citecolor=red}

\DeclareMathAlphabet\mathbfcal{OMS}{cmsy}{b}{n}

\def\beq{\begin{eqnarray}}\def\eeq{\end{eqnarray}}
\def\be{\begin{equation}}\def\ee{\end{equation}}

\def\a{\alpha}
\def\e{\epsilon}

\def\d{\delta}

\def\D{\Delta}
\def\G{\Gamma}
\def\l{\lambda}
\def\pd{\partial}

\def\Dphi{\Delta_{\phi}}
\def\bz{\bar{z}}
\def\la{\langle}
\def\ra{\rangle}

\def\G{\Gamma}

\def\bfAS{\bf\color{red}}

\def\beq{\begin{eqnarray}}\def\eeq{\end{eqnarray}}
\def\be{\begin{equation}}\def\ee{\end{equation}}

\def\a{\alpha}
\def\e{\epsilon}

\def\d{\delta}

\def\D{\Delta}
\def\G{\Gamma}
\def\l{\lambda}
\def\pd{\partial}

\def\bz{\bar{z}}
\def\la{\langle}
\def\ra{\rangle}

\def\G{\Gamma}

\newenvironment{comment}{\begingroup\color{blue}\itshape}{\endgroup}
\usepackage{bm}
\usepackage{bm}

\begin{document}

\title{\bf Positive geometry in the diagonal limit of the conformal bootstrap}
\date{}

\author{\!\!\!\! Kallol Sen$^{a}$\footnote{kallolmax@gmail.com}, Aninda Sinha$^{b}$\footnote{asinha@iisc.ac.in} and Ahmadullah Zahed$^{b}$\footnote{ahmadullah@iisc.ac.in}\\ ~~~~\\
\it ${^a}$Kavli Institute for the Physics and Mathematics of the Universe (WPI),\\
\it University of Tokyo, Kashiwa, Chiba 277-8583, Japan\\ \\
\it ${^b}$Centre for High Energy Physics,
\it Indian Institute of Science,\\ \it C.V. Raman Avenue, Bangalore 560012, India. }
\maketitle
\maketitle
\vskip 2cm
\abstract{We consider the diagonal limit of the conformal bootstrap in arbitrary dimensions and investigate the question if physical theories are given in terms of cyclic polytopes. Recently, it has been pointed out that in $d=1$, the geometric understanding of the bootstrap equations for unitary theories leads to cyclic polytopes for which the faces can all be written down and, in principle, the intersection between the unitarity polytope and the crossing plane can be systematically explored. We find that in higher dimensions, the natural structure that emerges, due to the inclusion of spin, is the weighted Minkowski sum of cyclic polytopes. While it can be explicitly shown that for physical theories, the weighted Minkowski sum of cyclic polytopes is not a cyclic polytope, it also turns out that in the large conformal dimension limit it is indeed a cyclic polytope. We write down several analytic formulae in this limit and show that remarkably, in many cases, this works out to be very good approximation even for $O(1)$ conformal dimensions. Furthermore, we initiate a comparison between usual numerics obtained using linear programming and what arises from positive geometry considerations.}

\tableofcontents

\onehalfspacing

\section{Introduction}
In a CFT, the four point function of identical scalars with dimension $\Dphi$ can be written as sum over conformal blocks of exchange operator $\mathcal{O}_{\D,\ell}$ of scaling dimension $\D$, spin $\ell$ with OPE coefficient (squared) $C_{\D,\ell}$:
\be
\left\langle\phi\left(x_{1}\right) \phi\left(x_{2}\right) \phi\left(x_{3}\right) \phi\left(x_{4}\right)\right\rangle=\frac{1}{\left|x_{12}\right|^{2 \Delta_{\phi}}\left|x_{34}\right|^{2 \Delta_{\phi}}} \mathcal{A}(u,v)
\ee
with
\be
\mathcal{A}(u,v)=\sum_{\D,\ell}C_{\D,\ell} G_{d,\D,\ell}(u,v)\,,
\ee
and $C_{\D,\ell}>0$ for unitary theories. Crossing symmetry implies
\be
\mathcal{A}(u,v)=(\frac{u}{v})^{\Dphi} \mathcal{A}(v,u)\,.
\ee
 This is the usual starting point for modern studies of the conformal bootstrap. One can consider the diagonal limit of this equation which is obtained by setting $u=z\bar z, v=(1-z)(1-\bar z)$ and taking the limit $z\rightarrow \bar z$. Thereby we obtain a problem in 1-variable. Following the numerical approach based on linear programming, pioneered in \cite{rrtv, rych, stuff} and recently reviewed in \cite{prv}, one can study this equation in any dimensions. Somewhat surprisingly\footnote{This feature appears to have gone unnoticed in the literature. One can do the analogous exercise in other dimensions. For $d=1$, there is a kink at $\Dphi=0$ which is within the unitarity limit but for $d=3$ it is at $\Dphi\sim 0.26$ and for $d=4$ it is at $\Dphi\sim 0.5$ both of which are below the unitarity bound.}, for $d=2$, retaining only scalar operators one finds a ``kink'' in the plot of the allowed leading scalar operator dimension when plotted against $\Dphi$. In figure 1, we give a comparison between the numerics obtained in the diagonal limit and what is obtained in the full problem\footnote{We thank Slava Rychkov for generously sharing the notebook for the full problem solved in \cite{rychkovvichi}. In the approach of \cite{rrtv}, one writes $z=1/2+a+b,\bar z=1/2+a-b$. If one suppresses the $b$ derivatives in the basis, then one will recover the diagonal limit numerics in fig. 1.}. The kink in the diagonal problem is at $\Dphi\approx 0.09, \D_2\approx 1.2$ and adding other operators with spin does not appear to change the location drastically (although we have not attempted to make a thorough comparison). The full problem \cite{rychkovvichi} in which there are off-diagonal derivatives,  needs the inclusion of higher spin blocks before a kink appears. Furthermore, compared to the full problem where the kink appears at $\Dphi\approx 0.125, \D_2\approx 1$, the kink in the diagonal limit is displaced. Nonetheless, one can ask what causes the kink in this instance. In the full problem as reviewd in \cite{prv}, the kink arises due to decoupling of certain operators (which happens for physical theories like the 2d Ising model due to null state conditions). 
\begin{figure}[h]\begin{center}
    \includegraphics[width=0.5\linewidth]{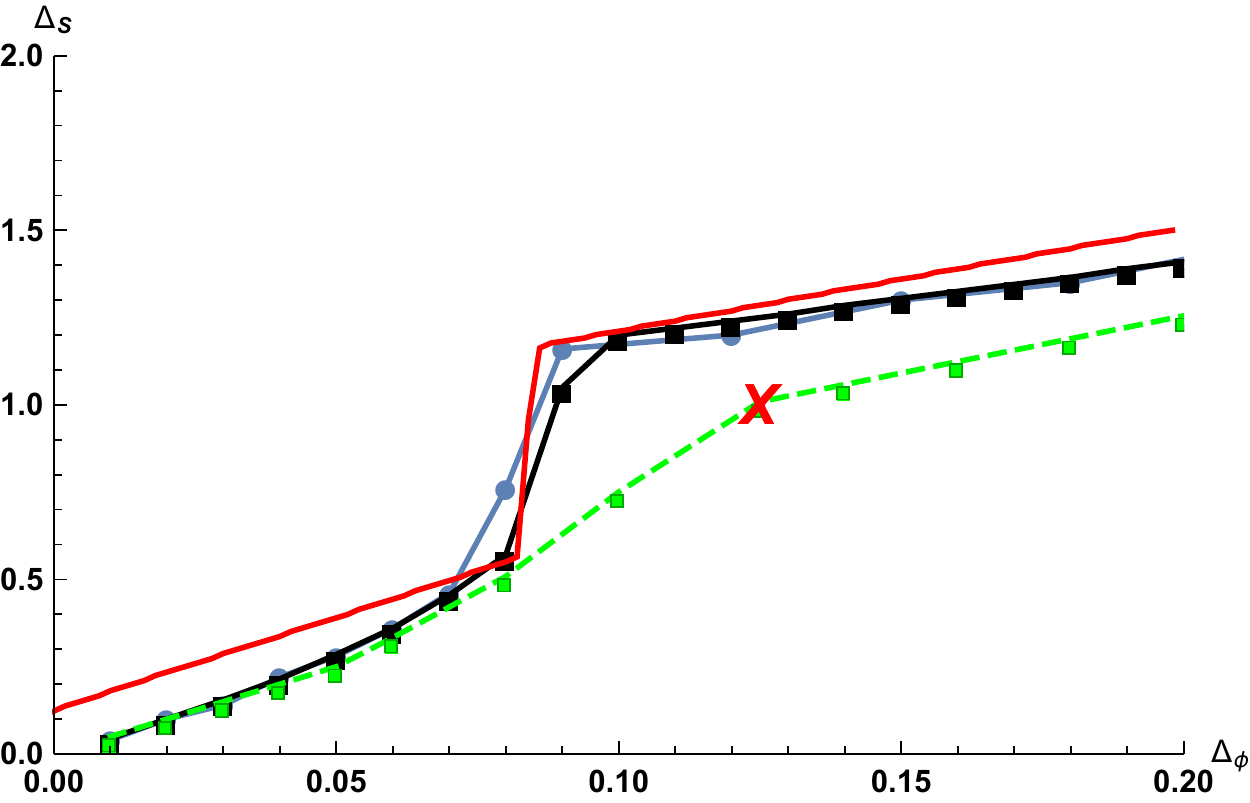}
    \caption{Plot of bounds on dimension of leading scalar in the OPE of $\phi\times \phi$ for $d=2$ using linear programming, keeping only scalars in the spectrum in the diagonal limit. The blue line is for 10 derivatives in the functional while the black is for 20 derivatives. In each case truncating the spectrum does not change the plot much. The green line is obtained in the non-diagonal limit \cite{rychkovvichi}, including spins, while the red cross is the location of the 2d Ising model. The red line indicates what arises from considering the intersection of the $N=2$ cyclic polytope obtained using scalars in $d=2$ and the crossing plane.}\end{center}
  \end{figure}

Thus the diagonal limit (being somewhat simpler since it is a one variable problem) appears to be a good playground to investigate various features of the conformal bootstrap. Recently the one variable bootstrap in $d=1$ has been studied and a deep connection with the alternative approach to the bootstrap advocated in \cite{polya, pap1,pap2,pap3,pap4,pap5,gs} and based on crossing symmetric Witten diagram basis expansion \cite{pap2, gs}, has been found \cite{Mazac:2018mdx}. Another development that has happened last year is an attempt to connect the usual conformal bootstrap program to the positive geometry of cyclic polytopes \cite{nima}. In this paper, we will endeavour to further this connection.

The work of \cite{rrtv} already made use of the polytope structure underlying the bootstrap equations. The idea was to rewrite the bootstrap equations in the form 
\be\label{cross}
\sum_{\D,\ell}C_{\D,\ell}\mathcal{F}_{d,\D,\ell}(z,\bar z)=1\,,
\ee 
where $\mathcal{F}_{d,\D,\ell}$'s are certain linear combinations of the direct and crossed-channel blocks given by
\be
\ \mathcal{F}_{d,\D,\ell}(z,\bz)=\frac{v^{\D_\phi}G_{d,\D,\ell}(z,\bz)-u^{\D_\phi}G_{d,\D,\ell}(1-\bz,1-z)}{u^{\D_\phi}-v^{\D_\phi}}\,,
\ee
where $u=z\bz$ and $v=(1-z)(1-\bz)$.
 Then, taking derivatives of the above equation around $z,\bar z=\frac{1}{2}$, one assumes a certain spectrum where the lowest scalar has dimension $\D_S$ and asks if the crossing condition can be satisfied. Using linear programming, one can rule out potential spectra and obtain bounds on $\D_S$ amongst other quantities. The starting point in \cite{nima} is somewhat different. One first expands the correlator (in \cite{nima} it was done in $d=1$), in terms of the direct channel blocks. The key novel insight provided in \cite{nima} is that if $\D$ is not too small, then this expansion has the structure of a cyclic polytope. The reason why this is important is that the face structure of the cyclic polytope has been fully classified. This enables one to study the intersection of the crossing plane with the polytope systematically and obtain various non-trivial insights about a potential CFT satisfying the bootstrap constraints. 

In this paper we will extend the work of \cite{nima} in several directions. We will consider the diagonal limit of the bootstrap equations in arbitrary dimensions. Rather than an expansion in terms of SL(2,R) blocks, this entails an expansion in terms of the so-called diagonal blocks worked out in \cite{diag}. We will show that one can work out a systematic $1/\D$ expansion of the blocks\footnote{The efficacy of the $1/\D$ expansion has also been recently discussed in \cite{zhibo}.} which enables an analytic study of the results of \cite{nima}. Furthermore, the cyclic polytope picture in \cite{nima} enables us to initiate a comparison of how the kink in fig.1 is obtained in the approach of \cite{rrtv} and from the cyclic polytope perspective. Compared to the $d=1$ study in \cite{nima} where the sum over spectrum only involved the conformal dimension, here we will also have to sum over spin. This creates an interesting difficulty. While for each spin, there is a cyclic polytope structure, when we sum over spins we get what is called a weighted Minkowski sum of cyclic polytopes \cite{bai}. The resulting polytope is not necessarily a cyclic polytope and one can easily check that for known CFTs like the 2d/3d Ising models, it is not. For low number of vertices, one can of course grind out the face structure. But that would miss the point made in \cite{nima}. What we find however, is that in the large $\D$ limit the weighted Minkowski sum is again a cyclic polytope! This means that at least in this limit, we can begin understanding (even analytically) the bootstrap problem using the positive geometry of cyclic polytopes as in \cite{nima}.

This paper is organized as follows. In section 2, we review known results about the diagonal limit of conformal blocks in arbitrary dimensions and present formulae for the leading order large $\Delta$ limit of the blocks. In section 3, we review salient features of polytopes, cyclic polytopes and conditions for the intersection of the crossing plane with the polytope. In section 4, we investigate the positivity criteria for cyclic polytopes both numerically and analytically. In sections 5 and 6 we consider the $N=1$ and $N=2$ polytopes and present asymptotic formulae for the bounds in \cite{nima} as well as analogous results in higher dimensions. In section 7 we initiate a study of numerical bootstrap using the cyclic polytope picture but closer in spirit to the approach \cite{rrtv} which used linear programming. We conclude in section 8. Some appendices support various calculations in the main text.

We will reserve the notation $d$ for spacetime dimensions and $D$ for the dimensionality of the polytope.

\section{Diagonal limit of conformal blocks}
In order to use the technology presented in \cite{nima} in higher dimensions, one would like to consider the diagonal limit of higher dimensional conformal blocks. This was worked out in \cite{diag}. 
\subsection{Exact blocks}
Unlike $d=1$, where there are no spins, the higher dimensions the spectrum of operators are labelled by two quantum numbers: $\D$ and $\ell$ and hence a sum over these two quantum numbers\footnote{We consider the block for the even spin. For odd spin, which will be relevant for say $O(N)$ model, the expression is slightly different \cite{diag}.}, 
\be\label{block}
\begin{split}
G_{d,\D,\ell}(z)=&\frac{\left(\frac{z^2}{1-z}\right)^{\Delta /2} (d-2)_{\ell } \left(\frac{\Delta +1}{2}\right)_{\frac{\ell }{2}}}{\left(\frac{d-2}{2}\right)_{\ell } \left(\frac{\Delta }{2}\right)_{\frac{\ell }{2}} \left(\frac{1}{2} (-d-\ell +\Delta +3)\right)_{\frac{\ell }{2}}}\sum _{r=0}^{\frac{\ell }{2}} \frac{\left(\frac{1}{2}\right)_r \binom{\frac{\ell }{2}}{r} \left(\frac{d-2+\ell}{2}\right)_r \left(\frac{2-d+\D-\ell}{2}\right)_{\frac{\ell }{2}-r} ~}{\left(\frac{1 +\Delta }{2}\right)_r} \\
&\times \, _3F_2\left(-\frac{d}{2}+\frac{\Delta }{2}+1,r+\frac{\Delta }{2},\frac{\Delta }{2};r+\frac{\Delta }{2}+\frac{1}{2},-\frac{d}{2}+\Delta +1;\frac{z^2}{4 (z-1)}\right)\,.
\end{split}
\ee
The ${}_3F_2$ above can again be of course again decomposed in terms of a sum over the $SL(2,R)$ blocks. We will not however venture in that direction. We note specifically, for the scalar operators, 
\be\label{blockscal}
G_{d,\D,0}(z)=\left(\frac{z^2}{1-z}\right)^\frac{\D}{2}{}_3F_2\bigg[\begin{matrix}1+\frac{\D-d}{2}, \frac{\D}{2},\frac{\D}{2}\\ \frac{\D+1}{2},\D+1-\frac{d}{2}\end{matrix};\frac{z^2}{4(z-1)}\bigg]\,.
\ee
We plot the second derivative of $\mathcal{F}_{d,\D,\ell}$ as a function of $z$ in fig.2 for $d=2$. For $\Dphi\lesssim 0.35$, for non-zero spin, unitary operators, the second derivative is positive around $z=1/2$. For $\ell=0$, the second derivative is positive for $\D>\D_S$ for some $\D_S$ which necessitates there being at least one scalar operator of $\D\leq \D_S$ for the bootstrap conditions to be satisfied \cite{rrtv}. What happens beyond $\Dphi\gtrsim 0.35$ is that some non-zero spin unitary operators (beginning with $\ell=2$) have negative second derivatives. Therefore, it is in principle possible to satisfy the bootstrap constraints in this case where we include at least one operator with $\D_{\ell=2}\leq \D_T$ for some $\D_T$ without necessarily having a scalar $\D\leq \D_S$.
\begin{figure}
\begin{tabular}{cc}
\includegraphics[width=0.45\linewidth]{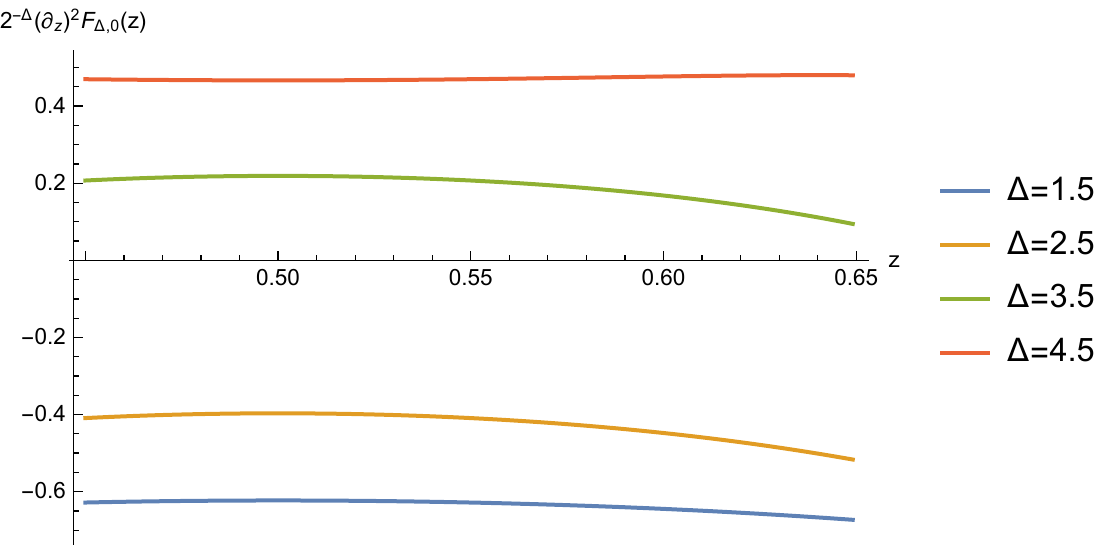} &\includegraphics[width=0.45\linewidth]{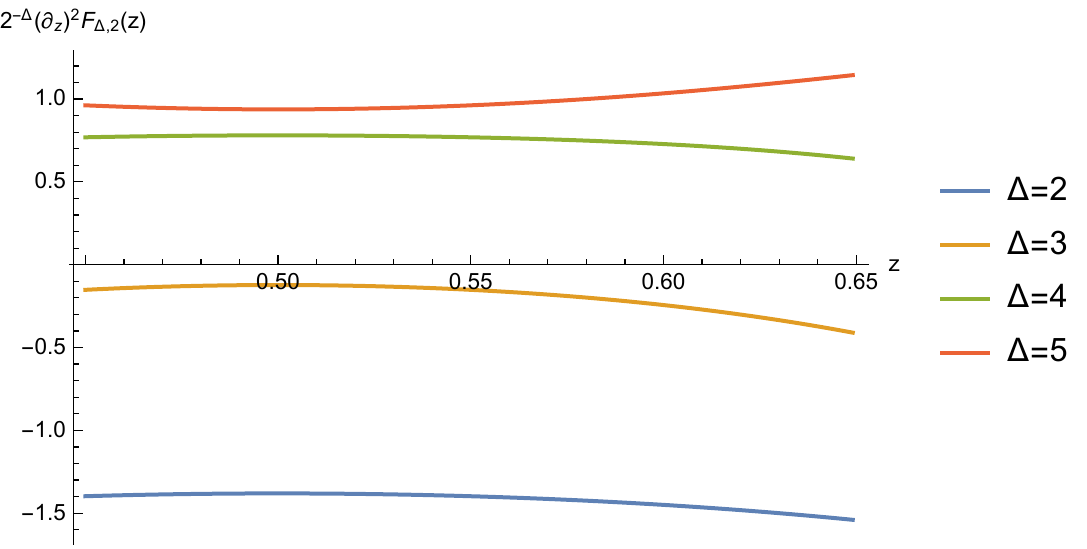}\\
(a) & (b)
\end{tabular}
\caption{Plot of $\partial_z^2 \mathcal{F}_{d,\D,\ell}(z)$ as a function of $z$ for $\Dphi=0.75$ in $d=2$. For $\ell\geq 4$ for unitary $\D\geq \ell$, the second derivative is positive in the region shown above. }
\end{figure}

\subsection{Asymptotic blocks}\label{asymbl}
 In this paper, at various stages we will use the large-$\D$ limit of the diagonal conformal blocks. The result we will need was worked out by G. N. Watson in 1914 and can be found in \cite{luke1}:
 \be
\begin{split}\label{3f2large}
&\, _3F_2\left(-\frac{d}{2}+\frac{\Delta }{2}+1,r+\frac{\Delta }{2},\frac{\Delta }{2};r+\frac{\Delta }{2}+\frac{1}{2},-\frac{d}{2}+\Delta +1;Z\right)\sim\\
&\frac{\left(\frac{1-\sqrt{1-Z}}{Z}\right)^{-\frac{d}{2}} \left(2^{\Delta -\frac{d}{2}} \left(-Z+2 \sqrt{1-Z}+2\right)^{-\frac{\Delta }{2}}\right)}{\sqrt{-Z-2 \sqrt{1-Z}+2}}\Bigg[\left(\sqrt{1-Z}-1\right)-\frac{\left(Z+2 \sqrt{1-Z}-2\right) \left(d^2-6 d-8 r+4\right)}{8 \Delta }\\
&+O\left(\frac{1}{\D^2}\right)\Bigg]
\end{split}
\ee
where $Z=\frac{z^2}{4 (z-1)}$. 
In terms\cite{diag} of $Z=-\frac{4 \rho^2}{(1-\rho^2)^2}$, or $z=\frac{4\rho}{(1+\rho)^2}$, with $0\leq \rho\leq 1$, this is very simply
$$(1-\rho^2)^{\D-\frac{d}{2}}\left[1+\frac{\left(d^2-6 d-8 r+4\right)}{8 \Delta }(\frac{2\rho^2}{1-\rho^2})+O\left(\frac{1}{\D^2}\right)\right]\,.$$
Under the crossing transformation $z\rightarrow 1-z$, $\rho\rightarrow \frac{(1-\sqrt{\rho})^2}{(1+\sqrt{\rho})^2}$.  This variable naturally arises in the large $\D$ asymptotics \cite{rrtv}. The blocks are analytic in the $z$ plane except for the cut $(1,\infty)$. $\rho$ maps the cut plane to a unit disc and can also be thought of as lightcone coordinates in higher dimensions \cite{caronhuot}. The general form is that the $1/\D^n$ term in the series inside the bracket comes with $p_n(\rho^2,r,d)/(1-\rho^2)^n$ where $p_n(\rho^2,r,d)$ is a polynomial in $\rho^2,r,d$ of degrees $n,n,2n$ respectively. Many subleading orders can be systematically worked out following\cite{luke1}.
An interesting thing to notice is that the sub-leading term for $r=0$ changes sign when $d>6$. We do not know if this is of any significance.

Now using eq.(\ref{3f2large}) in eq.(\ref{block}), we get 
\be
\begin{split}
G_{d,\D,\ell}(z)\sim&\frac{\sqrt{\pi } 2^{\Delta +\frac{1}{2}} \left(\sqrt{1-Z}+1\right)^{d/2} \left(\frac{Z}{Z+2 \sqrt{1-Z}-2}\right)^{-\frac{\Delta }{2}} \Gamma \left(-\frac{d}{2}+\Delta +1\right) (d-2)_{\ell } \left(\frac{\Delta +1}{2}\right)_{\frac{\ell }{2}}}{\sqrt{\Delta } \Gamma \left(\frac{\Delta -1}{2}\right) \Gamma \left(\frac{1}{2} (-d+\Delta +3)\right) \left(\frac{d-2}{2}\right)_{\ell } \left(\frac{\Delta }{2}\right)_{\frac{\ell }{2}}}\\
&\sum_{r=0}^{\ell/2}\frac{\left(\frac{1}{2}\right)_r \binom{\frac{\ell }{2}}{r} \left(\frac{1}{2} (d+\ell -2)\right)_r \left(\frac{1}{2} (-d-\ell +\Delta +2)\right)_{\frac{1}{2} (\ell -2 r)}}{\left(\frac{\Delta +1}{2}\right)_r \left(\frac{1}{2} (-d-\ell +\Delta +3)\right)_{\frac{\ell }{2}}}\left(1+O(\frac{1}{\D})\right)
\end{split}
\ee
where $Z= \frac{z^2}{4 (z-1)}$. Then doing the $r$ rum and using the Saalschutz formula 
\be
\, _3F_2\left(\frac{1}{2},\frac{d}{2}+\frac{\ell }{2}-1,-\frac{\ell }{2};\frac{d}{2}-\frac{\Delta }{2},\frac{\Delta }{2}+\frac{1}{2};1\right)=\frac{\left(\frac{\Delta }{2}\right)_{\ell/2 } \left(\frac{1}{2} (d-\Delta -1)\right)_{\ell/2 }}{\left(\frac{\Delta +1}{2}\right)_{\ell/2 } \left(\frac{d-\Delta }{2}\right)_{\ell/2 }}\,,
\ee
we finally get to the leading order, 
\be\label{blockapprox}
\begin{split}
G^{approx}_{d,\D,\ell}(z)\sim\frac{\sqrt{\pi} 2^{-\frac{3 d}{2}+2 \Delta +3} \left(\sqrt{1-Z}+1\right)^{d/2} \left(\frac{Z}{Z+2 \sqrt{1-Z}-2}\right)^{-\frac{\Delta }{2}} \Gamma (d+\ell -2)}{\Gamma \left(\frac{d-1}{2}\right) \Gamma \left(\frac{d}{2}+\ell -1\right)}[1+O(\D^{-1})]\,.
\end{split}
\ee
In terms the $\rho$ variable, we note 
$$
\left(\sqrt{1-Z}+1\right)^{d/2} \left(\frac{Z}{Z+2 \sqrt{1-Z}-2}\right)^{-\frac{\Delta }{2}}=2^{d/2} (1-\rho^2)^{-d/2}\rho^{\D}\,.
$$

\begin{figure}[ht]
  \centering
  \begin{subfigure}[b]{0.45\linewidth}
    \includegraphics[width=\linewidth]{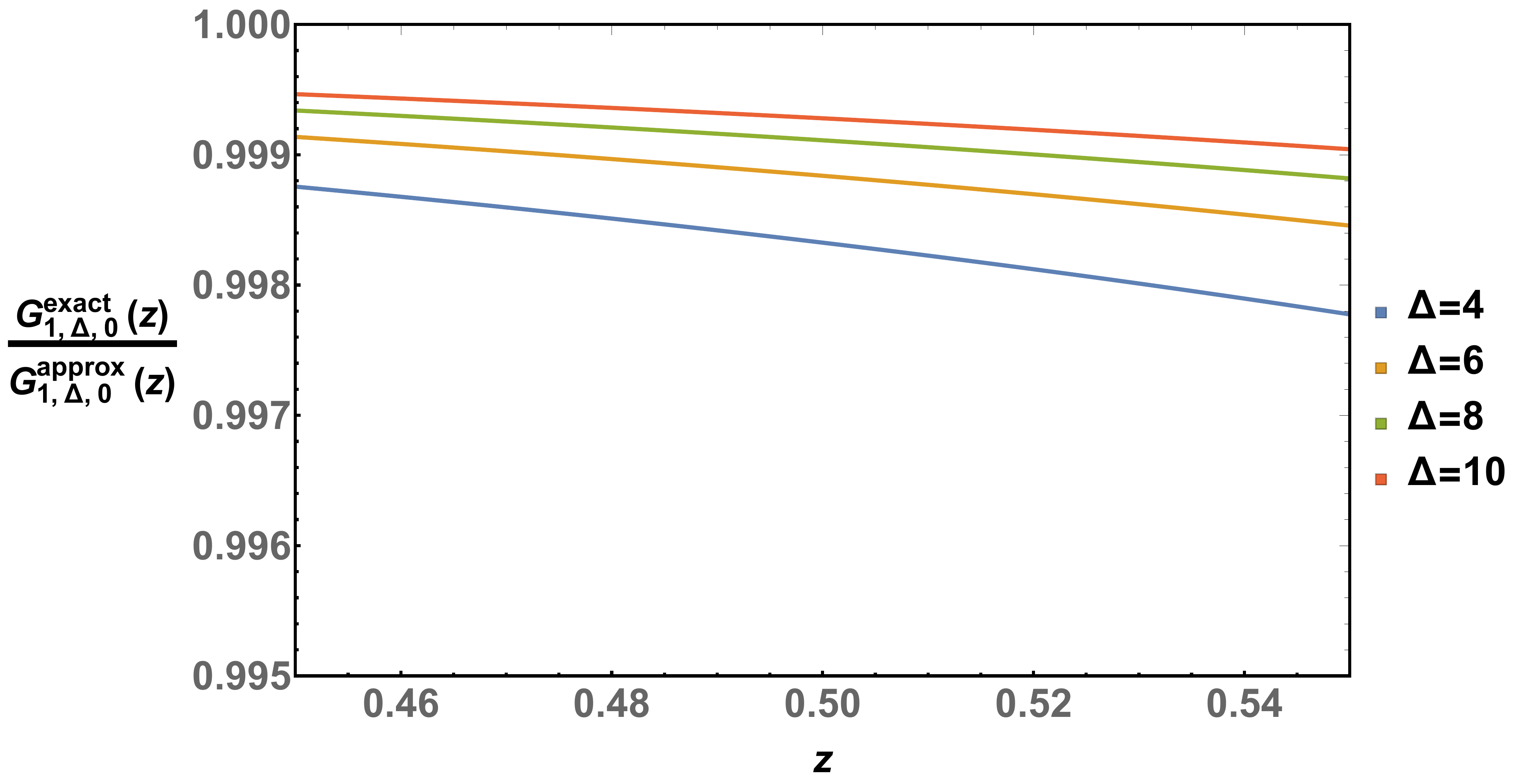}
    \caption{ Ratio of \ref{block} and \ref{blockapprox} vs $z$, $d=1$}
  \end{subfigure}
  \begin{subfigure}[b]{0.45\linewidth}
    \includegraphics[width=\linewidth]{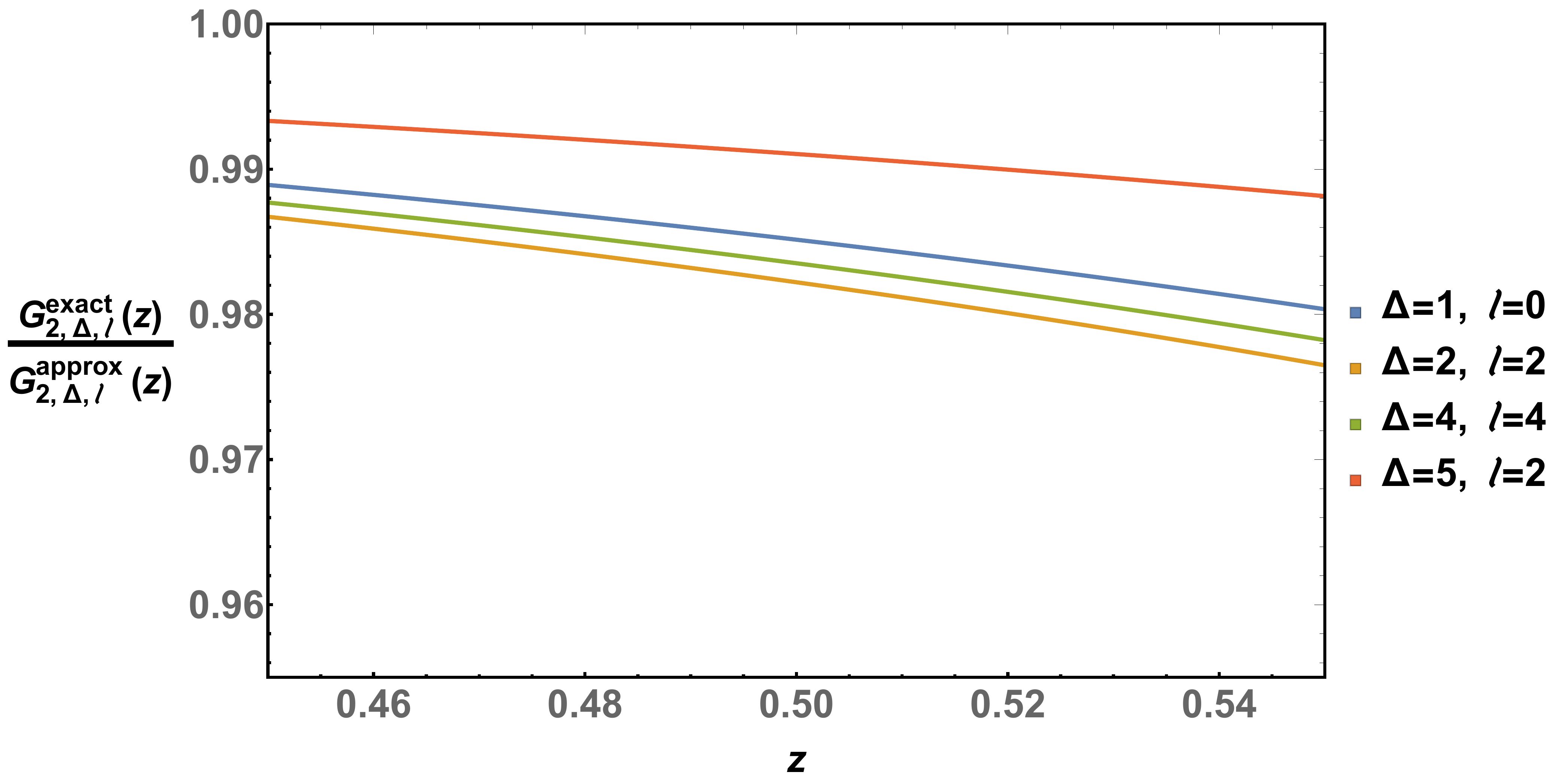}
    \caption{Ratio of \ref{block} and \ref{blockapprox} vs $z$, $d=2$ }
  \end{subfigure}
  \begin{subfigure}[b]{0.45\linewidth}
    \includegraphics[width=\linewidth]{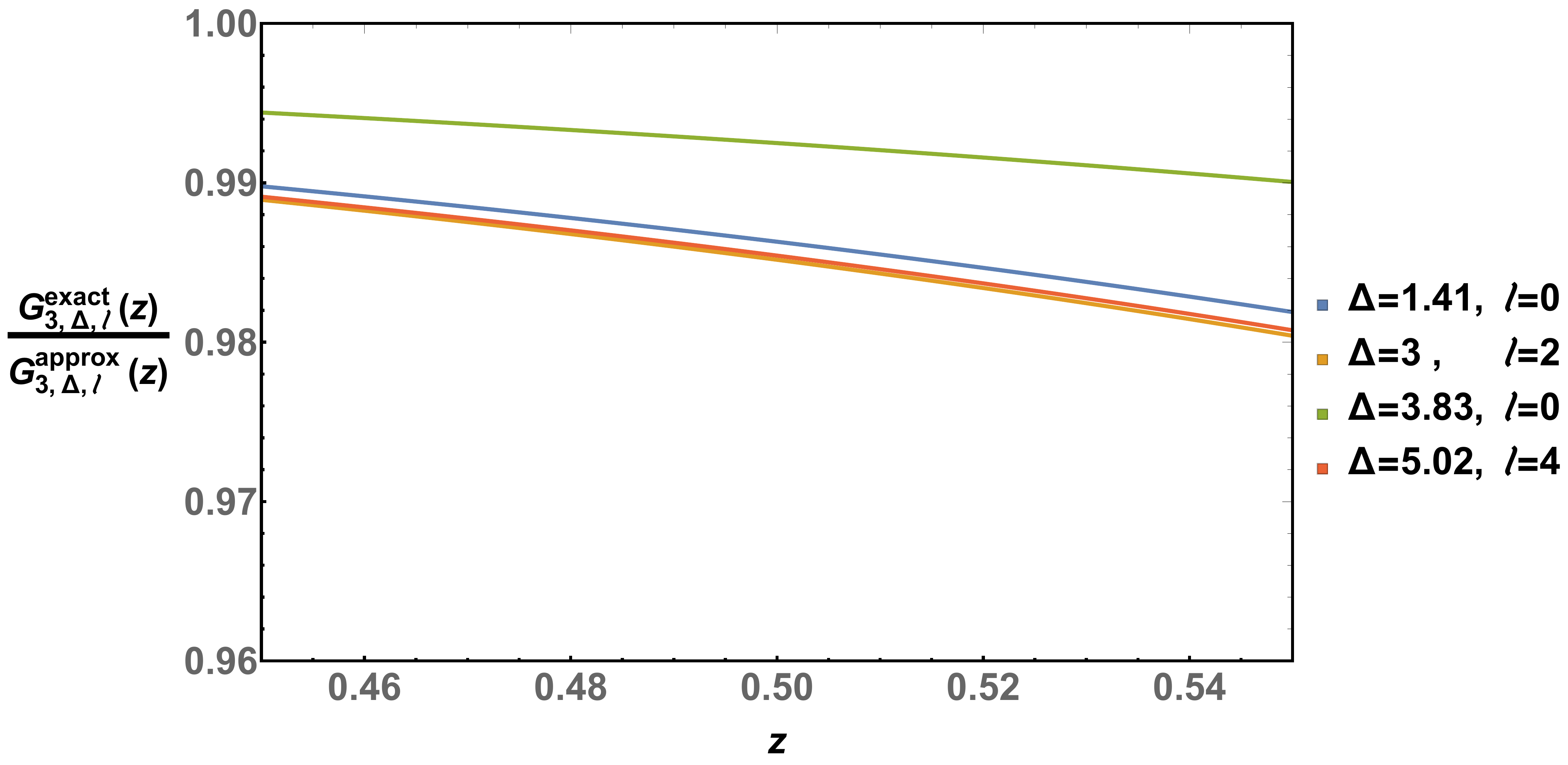}
    \caption{ Ratio of \ref{block} and \ref{blockapprox} vs $z$, $d=3$ }
  \end{subfigure}
  \caption{Ratio of \ref{block} and \ref{blockapprox} vs $z$ for various $\D$ and $\ell$. As is clear eq.(\ref{blockapprox}) gives an excellent approximation to the exact blocks for physical values of $\D,\ell$ to within a few percent. }
    \label{fig:ratio of block}
\end{figure}

We plot below the ratio of exact and asymptotic blocks. From fig.(\ref{fig:ratio of block}) it is clear that eq.(\ref{blockapprox}) is a reasonable approximation (upto $1\%$ error for even $\D\sim O(1)$). In many calculations below\footnote{We have worked out up to $1/\D^3$ terms in the asymptotic expansion but refrain from showing the hideous expressions in the main text.}, we will make use of these asymptotic blocks and derive analytical formulae which explain several numerical results arising from using the exact blocks.

\section{Polytopes: Unitary CFT and crossing equation}\label{cyclicpoly}

\subsection{Usual approach of CFT }

In this section, we will summarize certain key geometric ideas given in \cite{nima} as well as \cite{rrtv} which will be used in the paper.
To start with, we consider the geometric meaning of the sum rule presented in eq. (\ref{cross}) in the diagonal limit.
\be\label{daigcross}
\begin{split}
 \mathcal{F}_{d,\D,\ell}(z)=\frac{(1-z)^{2\Dphi}G_{d,\D,\ell}(z)-z^{2\Dphi}G_{d,\D,\ell}(1-z)}{z^{2\Dphi}-(1-z)^{2\Dphi}}\,,
\end{split}
\ee
{\it In any dimension $d$, given a span of vectors $\mathcal{F}_{d,\D,\ell}(z)$ for a spectrum $\{\D,\ell\}$, does eq.\eqref{cross} admit a solution with arbitrary positive definite coefficients $C_{\D,\ell}$?}
\\
The {\it rhs} of \eqref{cross} defines a convex cone $\mathcal{C}$ in the space of functions $\{\mathcal{F}_{d,\D,\ell}(z)\}$. In order to proceed it is necessary to parameterize the space of functions $\{\mathcal{F}_{d,\D,\ell}(z)\}$. This is achieved by replacing the functional space by a discrete infinite vector space of derivatives of $\mathcal{F}_{d,\D,\ell}(z)$ at some symmetric point. For convenience the most symmetric point (as also the most convergent point of expansion) is $z=1/2$. The vector space is made of even derivatives\footnote{We have borrowed notation from \cite{rrtv}},
\be
\mathcal{B}=\{\mathcal{F}_{d,\D,\ell}^{2m}\}=\bigg\{\pd_z^{2m} \mathcal{F}_{d,\D,\ell}(z)\bigg|_{z=1/2}\bigg\}\,.
\ee
while the odd derivatives vanish due to crossing symmetry in \eqref{daigcross}. In terms of the basis, \eqref{daigcross} can be written as a set of infinite relations for $m\geq0$,
\be\label{cross1}
\sum_{\D,\ell}C_{\D,\ell}\mathcal{F}^{2m}_{d,\D,\ell}=\begin{cases}1\,, m=0\\ 0\,, m>0\end{cases}\,.
\ee
The key idea is to identify the linear functional,
\be
\Lambda: \ \Lambda(\mathcal{F}_{d,\D,\ell})=\sum_{\mathcal{B}}\lambda_{2m}\mathcal{F}^{2m}_{d,\D,\ell}\,,
\ee
such that, \eqref{cross} can be checked via numerical methods such as {\it Linear Programming}. Refraining from the details which are explained nicely in \cite{rrtv}, the essence of the approach is:
\begin{itemize}
\item Assume a spectrum consistent with unitarity and assuming that scalars have dimension $\D\geq \D_S$. 
\item A solution to the linear programming problem gives a set of $\lambda_{2m}$ such that $\sum_{\D,\ell}C_{\D,\ell}\lambda_{2m}\mathcal{F}^{2m}_{\D,\ell}>0$. In other words for such a spectrum it is not possible to satisfy eq.\eqref{cross1} since this is nothing but a linear combination of the equations in eq.\eqref{cross1}. 
\item Reduce $\D_S$ such that no solution to the linear programming problem is found. This corresponds to a potentially allowed spectrum.
\end{itemize}

\subsection{Positive geometric approach of CFT }

The approach taken by \cite{nima} is somewhat different and makes use of the known faces of what is called a cyclic polytope. Instead of starting with eq.\ref{cross}, one can equivalently consider the convex cone $\mathcal{C}$ or more specifically the {\it convex hull}  of vectors built from the conformal blocks themselves.\\
Specifically, one truncates the Taylor series of the four-point function, which in turn is expanded in terms of the direct channel conformal blocks, upto some say $2N+2$ terms around $z=1/2$
\be
\begin{split}
\mathcal{A}(z)=&\mathcal{A}^{0}+\mathcal{A}^{1} y+\mathcal{A}^{2} y^{2}+\cdots +\mathcal{A}^{2N+1}y^{2N+1}\\
G_{d,\D,\ell}(z)=&G_{d,\D,\ell}^{0}+G_{d,\D,\ell}^{1} y+G_{d,\D,\ell}^{2} y^{2}+\cdots +G_{d,\D,\ell}^{2N+1}y^{2N+1}
\end{split}
\ee
where $y \equiv z-1 / 2$ and $~\mathcal{A}^n=\frac{1}{n!}\pd_z^n\mathcal{A}(z)\bigg|_{z=1/2},~G_{d,\D,\ell}^n=\frac{1}{n!}\pd_z^n G_{d,\Delta,\ell}(z))\bigg|_{z=1/2}\,. $ Therefore $\mathbfcal{A}$ lies in the positive span of the block vector $\mathbf{G}_{d,\Delta,\ell} $  
\be\label{4p1}
\begin{split}
\mathbfcal{A}=\sum_{\D,\ell}C_{\D,\ell}\mathbf{G}_{d,\D,\ell}\,,
\end{split}
\ee
where $C_{\D,\ell}>0$  for unitary theories and,
\be
\mathcal{A}(z) \rightarrow \mathbfcal{A}=\left( \begin{array}{c}{\mathcal{A}^{0}} \\ {\mathcal{A}^{1}} \\ {\vdots} \\ {\mathcal{A}^{2 N+1}}\end{array}\right) \in \mathbb{P}^{2 N+1}~\text{and}~G_{d,\Delta,\ell}(z) \rightarrow \mathbf{G}_{d,\Delta,\ell}=\left( \begin{array}{c}{G_{d,\Delta,\ell}^{0}} \\ {G_{d,\Delta,\ell}^{1}} \\ {\vdots} \\ {G_{d,\Delta,\ell}^{2 N+1}}\end{array}\right) \in \mathbb{P}^{2 N+1}~\,.
\ee

The dimensionality $D$ of the polytope is $D=2N+1$. To make  $ \mathbb{P}^{2 N+1}$ explicit, we note that the above equation \ref{4p1} tells us that since we can trivially multiply both sides by a non-zero number (absorbing this into the $C_{\D,\ell}$'s on the rhs), we can equivalently consider the expansion of $\mathbfcal{A}~ \to t ~ \mathbfcal{A}$. Similarly the cone spanned by $\mathbf{G}_{d,\Delta,\ell}$ is same as $\mathbf{G}_{d,\Delta,\ell} \to \a_{\D,\ell} \mathbf{G}_{d,\Delta,\ell}$ with $\alpha_{\D,\ell}>0$ since we can again absorb this into the OPE coefficients. In other words we can expand   
\be
\mathbfcal{A}=t\begin{pmatrix}1\\ \vec{\mathcal{A}}\end{pmatrix}\,{\rm in~terms~of~} \mathbf{G}_{d,\Delta,\ell}=\a_{\D,\ell}\begin{pmatrix}1\\ \vec{G}_{d,\D,\ell}\end{pmatrix}\,, \ \ \,.
\ee
Now plugging back in to the equation \ref{4p1} gives $\sum \a_{\D,\ell}C_{\D,\ell}\equiv\sum C_{\D,\ell}'=t$. Therefore we can write the equation \ref{4p1} as 
\be\label{proj4p}
\begin{split}
\vec{\mathcal{A}}=\sum_{\D,\ell}\l_{\D,\ell}\vec{G}_{d,\D,\ell}\,,
\end{split}
\ee
where $\l_{\D,\ell}=\frac{ C_{\D,\ell}'}{\sum C_{\D,\ell}'}$. Also note that $\sum_{\D,\ell}\l_{\D,\ell}=1$ which defines convex hull of $\vec{G}_{d,\D,\ell}$ which is a polytope in $\mathbb{R}^{2 N+1}$. Therefore more generally we can think of the equation (\ref{4p1}) as projective polytope in  $ \mathbb{P}^{2 N+1}$.

It is also possible to define the polytope by a collection of hyperplanes $W_{aI}$, the facets, that cut out the polytope by the linear inequalities. Here $\mathbf{W}_{a}$ defines a collection of hyperplanes and
\be
\mathbfcal{A}\cdot \mathbf{W}_a=\mathcal{A}^I W_{aI}>0\,,
\ee
outline the structure of the polytope. It is difficult to check whether a given vector lies inside or outside a polytope. This is because the representation of $\mathbfcal{A}$ is non-unique. By contrast the facet definition does allow us to practically check whether or not a given $\mathbfcal{A}$ is inside or not by simply checking the inequalities.

Now given a  polytopes in $\mathbb{P}^{D}$, the facets for the convex hull of a set of vectors $\textbf{v}_i$  \footnote{$ \mathbf{v}_i$ or simply `` $i$ " in determinants are short hand notation for the components of $\mathbf{G}_{d,\D_i,\ell_i}$ } are given by the set of $D$ vectors $\left( \textbf{v}_{i_1},~\textbf{v}_{i_2},~\dots~\textbf{v}_{i_D} \right)$ such that \cite{nima},

\be\label{eq:def_sign}
\left\langle \textbf{v}_i,~\textbf{v}_{i_1},~\textbf{v}_{i_2} ~\dots~\textbf{v}_{i_D} \right\rangle ~\text{have same sign}~ ; ~ \forall~ i\,.
\ee
The corresponding facets defining the boundary of the polytope are,
\be
W_{aI}=\e_{I I_1\dots I_D}v_{i_1}^{I_1}\dots v_{i_D}^{I_D}\,.
\ee
 There is a class of polytope which vertices have an ordering $v_1,\dots v_n$ where $n$ can be larger than D, such that \cite{nima}. 
\be
\la \mathbf{v}_{i_1},\mathbf{v}_{i_2},\dots,\mathbf{v}_{i_D}\ra\,, \ \text{have same sign}\ \forall\ i_1<i_2<\dots<i_D\,.
\ee
Such structures can be analytically computed for any $\mathbb{P}^D$ and form the {\it cyclic polytope}. The facets of a cyclic polytope are given by \cite{nima}

\be\label{eq:cyclic_face}
\begin{split}
\text{D odd: }~\left\{\mathbf{W}_{a}\right\}=&\left(\mathbf{v}_{1}, \mathbf{v}_{i_{1}}, \mathbf{v}_{i_{1}+1}, \mathbf{v}_{i_{2}}, \mathbf{v}_{i_{2}+1}, \cdots, \mathbf{v}_{i_{\frac{D-1}{2}}}, \mathbf{v}_{i_{\frac{D-1}{2}}+1}\right)\\
&\cup(-1)\left(\mathbf{v}_{i_{1}}, \mathbf{v}_{i_{1}+1}, \mathbf{v}_{i_{2}}, \mathbf{v}_{i_{2}+1}, \cdots, \mathbf{v}_{i_{\frac{D}{2}}-1}, \mathbf{v}_{i _{\frac{D-1}{2}}+1}, \mathbf{v}_{n}\right)\,,\\
\text{D even: }~\left\{\mathbf{W}_{a}\right\}=&\left(\mathbf{v}_{i_{1}}, \mathbf{v}_{i_{1}+1}, \mathbf{v}_{i_{2}}, \mathbf{v}_{i_{2}+1}, \cdots, \mathbf{v}_{i_{\frac{d}{2}}}, \mathbf{v}_{i_{\frac{d}{2}}+1}\right)\,.
\end{split}
\ee

\subsubsection{Positivity of the blocks}\label{sec:gi}

 From the point of view of the CFT spectrum, the block vectors can be ordered simply in terms of increasing $\D$. This would mean that the spins are not necessarily ordered.  Additionally there are two more vectors $\mathbf{v}_0=(1,0,\dots,0)$ and $\mathbf{v}_\infty=(0,\dots,0,1)$ which represent the identity operator and the $\D=\infty$ operator as a part of the spectrum. In order to show that block vectors $\mathbf{G}_{d,\D,\ell}$ form a cyclic polytope we proceed as below: First we have an ordered set of vectors $(i_i, i_2,\cdots i_{D+1})$ such that $\D_{i_1}<\D_{i_2}<\cdots<\D_{i_{D+1}}$. Then in terms of the condition for positivity discussed in \cite{nima}, the condition for a cyclic polytope is
\be\label{stability}
\left\langle i_{1}, i_{2}, \cdots, i_{D+1}\right\rangle \equiv \epsilon_{I_{1} I_{2} \cdots I_{D+1}} G_{d,\Delta_{i_{1}},\ell}^{I_{1}} \cdots G_{d,\Delta_{i_{D+1}},\ell}^{I_{D+1}} \  \text{have the same sign}\,.
\ee
In order to proceed from here we define, 
\be\label{Fmn}
F_{m,n}=\frac{1}{m!}~\partial _{\Delta }{}^n~\partial _z{}^m G_{d,\Delta ,\ell}(z)|_{z=1/2}\,.
\ee
Then we construct the $\mathbf{K}_{2N+1}$, $(2N+1)\times(2N+1)$ matrix,
\be
\mathbf{K}_{2N+1}(d,\D,\ell)= \left( \begin{array}{ccccc}{F_{0,0}} & {F_{1,0}} & { . .} & { . .} & {F_{2 N+1,0}} \\ {F_{0,1}} & {F_{1,1}} & { . .} & {. . } & { . .} \\ { . .} & { . .} & {F_{i, j}} & { . .} & { . .} \\ { . .} & { . .} & { . .} & { . .} & { . .} \\ {F_{0,2 N+1}} & { . .} & { . .} & { . .} & {F_{2 N+1,2N+1} }\end{array}\right)\,.
\ee
We can show that the condition for positivity discussed in \cite{nima}, can be written in a more generic format, given by\footnote{We have chosen the sign to be positive to concur with \cite{nima}.},
\be\label{gi}
g_{i}=\frac{\left|\mathbf{K}_{i}(d,\D,\ell)\right|\left|\mathbf{K}_{i-2}(d,\D,\ell)\right|}{\left|\mathbf{K}_{i-1}(d,\D,\ell)\right|^{2}}>0\,,
\ee
where $|\mathbf{K}_i|$ gives the determinant of a $i\times i$ matrix and $|\mathbf{K}_i|=1$ for $i<0$. With this, as we have checked rigorously, all the positivity criterion discussed in \cite{nima} can be reproduced {\it viz.} 
\be
\begin{split}
&N=0 : g_{1}=\left(G_{d,\Delta,\ell}^{1}\right)^{\prime}>0,\\
&N=1 : g_{2}=\left(\frac{\left(G_{d,\Delta,\ell}^{2}\right)^{\prime}}{\left(G_{d,\Delta,\ell}^{1}\right)^{\prime}}\right)^{\prime}>0,\quad g_{3}=\left(\frac{\left(\frac{\left(G_{d,\Delta,\ell}^{3}\right)^{\prime}}{\left(G_{d,\Delta,\ell}^{1}\right)^{\prime}}\right)^{\prime}}{\left(\frac{\left(G_{d,\Delta,\ell}^{2}\right)^{\prime}}{\left(G_{d,\Delta,\ell}^{1}\right)^{\prime}}\right)^{\prime}}\right)^{\prime}>0
\end{split}
\ee
and so on.

\subsection{The crossing equation}

So far we have discussed the notion of cyclic polytopes which is relatively simpler to compute analytically and associated with the notion of a collection of block vectors given in equation (\ref{4p1}). However in the numerical bootstrap, the convex hull of vectors is defined with respect to the crossing equation and not directly with the blocks themselves. In order to connect the concept of {\it cyclic polytope} to bootstrap, we will need to put in the ingredient of ``crossing" in the projective picture as well. Crossing in the diagonal limit reads
\be\label{cross_diag}
\mathcal{A}(z)=\left(\frac{z}{1-z}\right)^{2\D_\phi}\mathcal{A}(1-z)\,.
\ee

This relates odd $\mathcal{A}^{2n+1}$ in terms of the even ordered derivatives $\mathcal{A}^{2n}$. This in turn defines a hyperplane $\mathbf{X}[\D_\phi]$ which is a $(2N+2)\times(N+1)$ matrix in $\mathbb{P}^{2N+1}$. For example \cite{nima}, for $N=2$ or $D=2N+1=5$, 
\renewcommand{\kbldelim}{(}
\renewcommand{\kbrdelim}{)}
\be\label{N=2,X}
\begin{split}
\mathbf{\mathcal{A}}=&\left( \begin{array}{c}{\mathcal{A}^{0}} \\ {4 \Delta_{\phi} \mathcal{A}^{0}} \\ {\mathcal{A}^{2}} \\ {\frac{16}{3}\left(\Delta_{\phi}-4 \Delta_{\phi}^{3}\right) \mathcal{A}^{0}+4 \Delta_{\phi} \mathcal{A}^{2}} \\ {\mathcal{A}^{4}}\\ {\frac{64}{15} \Delta_{\phi}\left(32 \Delta_{\phi}^{4}-20 \Delta_{\phi}^{2}+3\right) \mathcal{A}^{0}-\frac{16}{3} \Delta_{\phi}\left(4 \Delta_{\phi}^{2}-1\right) \mathcal{A}^{2}+4 \Delta_{\phi} \mathcal{A}^{4}} \\ {\vdots}\end{array}\right) \in \mathbb{P}^{2 N+1}\\
\text{~~~~~~~~~~and}\\
\mathbf{X}[\Dphi]=& \kbordermatrix{
    & \mathcal{A}^0 &  \mathcal{A}^2 & \mathcal{A}^4 \\
     & 1 & 0  &0 \\
     & 4\Delta_\phi & 0 &0  \\
     & 0 & 1&0\\
     &{16\over3} (\Delta_\phi-4\Delta_\phi^3) & 4\Delta_\phi&0\\
     & 0 & 0&1\\
     & \frac{64}{15}\Delta_\phi(32\Delta^4_\phi-20\Delta_\phi^2+3) & {16\over3} (\Delta_\phi-4\Delta_\phi^3)&4\Delta_\phi
  }\,.
\end{split}
\ee

Now we have both ingredients for bootstrap in the projective picture. 
\begin{itemize}
\item A convex hull of vectors built from $2N+2$ components of a block with a natural notion of ordering. This defines a {\it cyclic polytope} in $\mathbb{P}^{2N+1}$. Call this $\mathbf{U}[\D]$.
\item A $N$-dimensional hyperplane $\mathbf{X}[\D_\phi]$ . 
\end{itemize}
The consistent solution of bootstrap entails finding $\mathbf{U}[\D]\cap \mathbf{X}[\D_\phi]$. The question now is given a $\mathbf{U}[\D]$, what are the conditions needed for the intersection with $\mathbf{X}[\D_\phi]$. As argued in \cite{nima}, a $k-$plane intersects with a $D-$dimensional polytope with a $D-k$ face at a point given by,
\be
\mathbf{v}_1\la \mathbf{v}_2, \mathbf{v}_3,\mathbf{v}_4,\dots,\mathbf{v}_{D-k},\mathbf{X}\ra-\mathbf{v}_2\la \mathbf{v}_1, \mathbf{v}_3,\mathbf{v}_4,\dots,\mathbf{v}_{D-k},\mathbf{X}\ra+\dots\,.
\ee
For a point inside the polytope and satisfying the intersection property above,
\be\label{eq:intsec}
\la \mathbf{v}_2, \mathbf{v}_3,\mathbf{v}_4,\dots,\mathbf{v}_{D-k},\mathbf{X}\ra\,, -\la \mathbf{v}_1, \mathbf{v}_3,\mathbf{v}_4,\dots,\mathbf{v}_{D-k},\mathbf{X}\ra\,, \la \mathbf{v}_1, \mathbf{v}_2,\mathbf{v}_4,\dots,\mathbf{v}_{D-k},\mathbf{X}\ra,~ \dots\,,
~~\text{have same sign}
\ee

 A further simplification occurs when one of the vertex vectors is the identity operator $\mathbf{v}_0$ or the infinity operator $\mathbf{v}_\infty$ since this reduces the dimensionality of the problem. Another simplification is when projecting through specific planes obtained by performing a $GL(D+1)$ transformation, 
\be
\mathbf{X}'=\Lambda_{GL(D+1)}\cdot \mathbf{X}\,.
\ee
In terms of the intersection problem, $\la\mathbf{X}',\dots\ra$ reduces to a $D-k+1$ dimensional determinant, reducing the rank by $k-1$.

\section{Positivity criterion for $\mathbf{U}[\D_i]$}

Here we discuss the positivity criterion of  $\mathbf{U}[\D_i]$ in diagonal limit.
As mentioned in \eqref{gi}, a sufficient condition for positivity of the block vector is $g_i>0$. Note however, that $g_i<0$ does not necessarily mean that the cyclic polytope structure is absent; it simply means that in this case one needs to check \eqref{stability} directly.

\subsection{Scalar blocks}
We plot below numerical evaluation of $g_i$ for the scalar block.
\begin{figure}[tb]
  \centering
  \begin{subfigure}[b]{0.47\linewidth}
    \includegraphics[width=\linewidth]{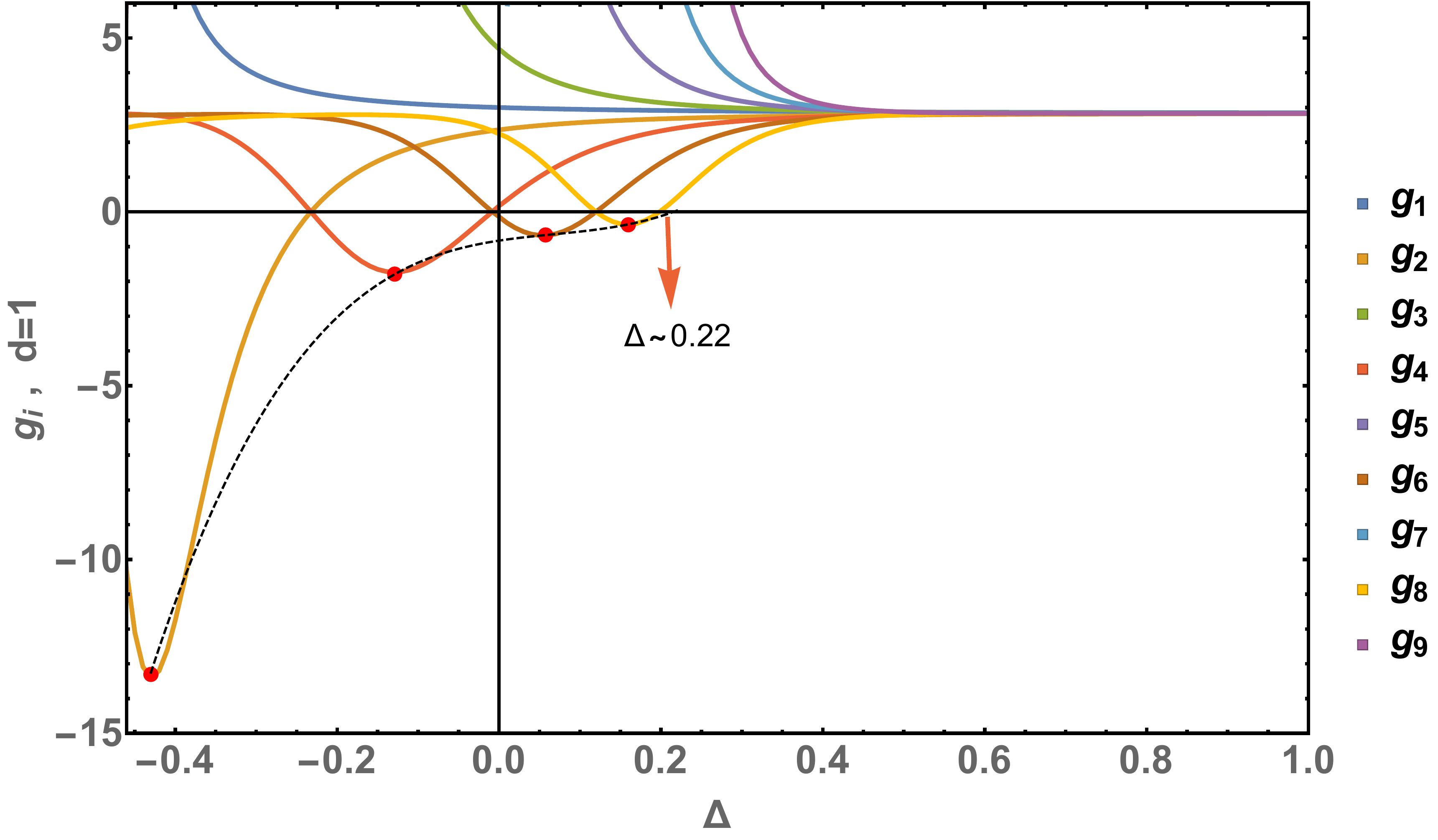}
    \caption{$g_i~vs~\D $ for $d=1$ }
  \end{subfigure}
  \begin{subfigure}[b]{0.47\linewidth}
    \includegraphics[width=\linewidth]{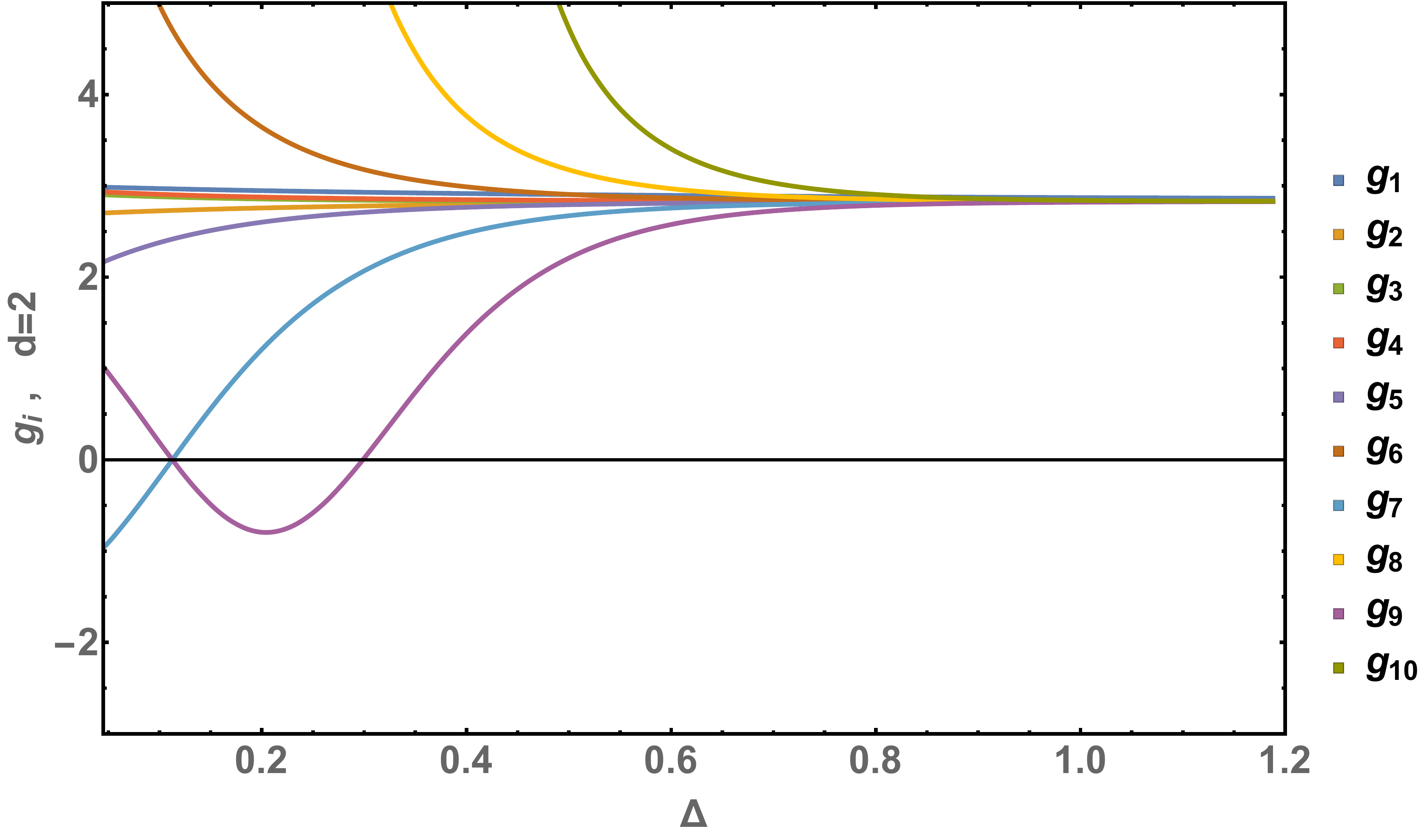}
    \caption{$g_i~vs~\D $ for $d=2$ }
  \end{subfigure}
  \begin{subfigure}[b]{0.47\linewidth}
    \includegraphics[width=\linewidth]{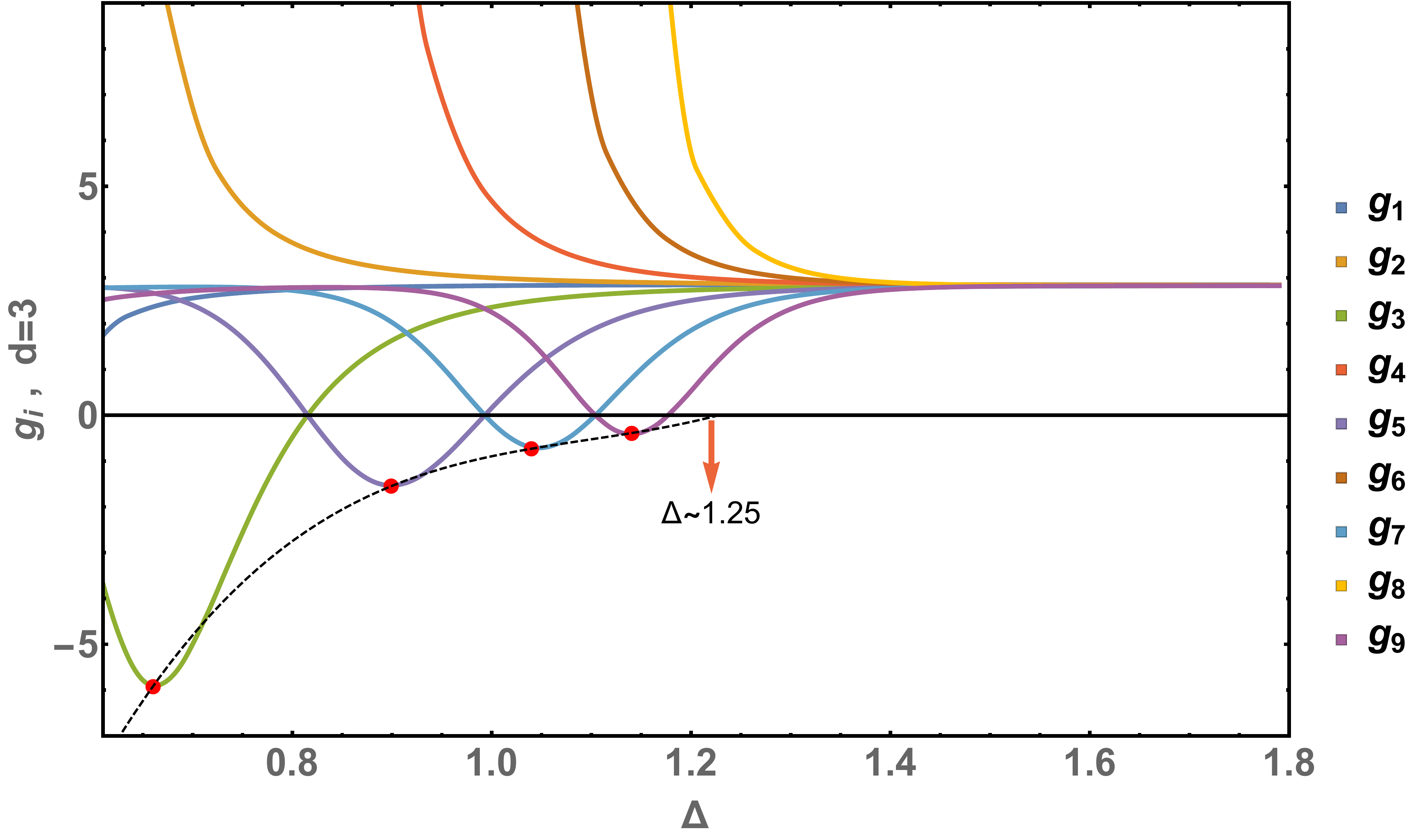}
    \caption{$g_i~vs~\D $ for $d=3$ }
  \end{subfigure}
  \begin{subfigure}[b]{0.47\linewidth}
    \includegraphics[width=\linewidth]{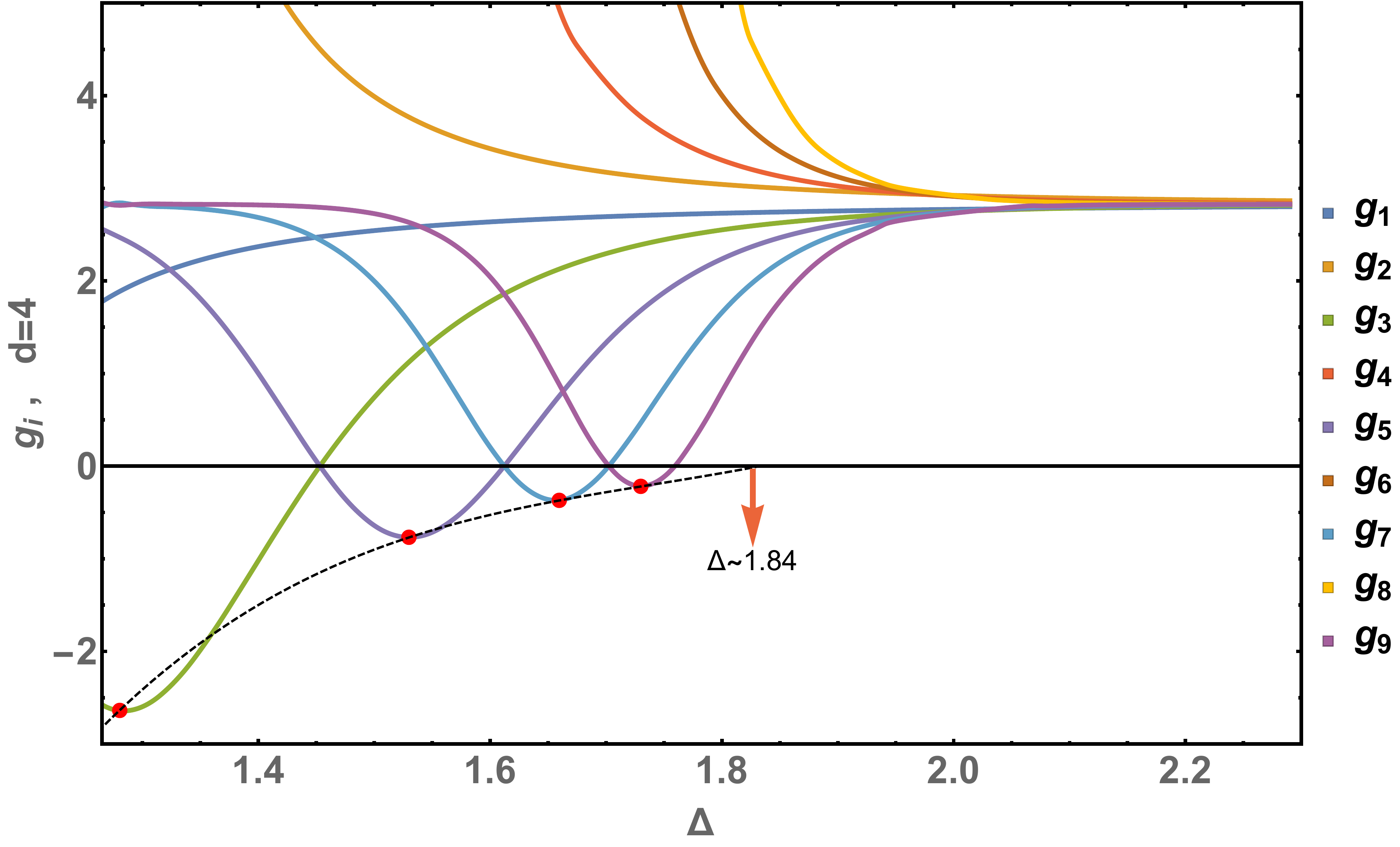}
    \caption{$g_i~vs~\D $ for $d=4$ }
  \end{subfigure}
  \caption{$g_i~vs~\D $ for scalar blocks for various $d$, plot range is $\D>\frac{d-2}{2}$. For $d=1,3,4$ a trend is exhibited where it can be expected that eventually all $g_i$'s become positive for some $\D$. For $d=2$ such a trend does not exist and one is forced to check eq.\eqref{stability} directly.  }
  \label{fig:scalarblock}
\end{figure}
From figure [\ref{fig:scalarblock}] it is clear that:
\begin{itemize}
\item $g_i\rightarrow 2\sqrt{2}$ $\forall\ i$ in the $\D\gg\frac{d-2}{2}$ limit.
\item Some of the $g_i$'s are negative around $\D\sim\frac{d-2}{2}$.
\end{itemize}
The minima of the negative $g_i$'s shift upwards for increasing $i$ \footnote{For $d=2$ we have not noticed any feature like this. }. From $\D\gg~ \frac{d-2}{2},\ell$ analysis we can track the $\D$ for which the minimum goes above the axis, which is indicated approximately in figure \ref{fig:scalarblock}.
\subsection{$\D\gg \frac{d-2}{2},\ell$ limit anaytic results}
We can explain some of the features of $g_i$ analytically by using the asymptotic blocks in section \ref{asymbl}.
For the scalar block we can parametrize the large $\Delta$ block as:
\be
G_{d,\D,0}^{approx}(z)=e^{A(z)\D}B(z)\left(1+\frac{c_1(z)}{\D}+\frac{c_2(z)}{\D^2}+O(\frac{1}{\D^3})\right)\,.
\ee
Using this and after coaxing mathematica a bit, we find
\be
g_n=A'(\frac{1}{2})-\frac{c_1'(\frac{1}{2})}{\D^2}+2\frac{c_1(\frac{1}{2})c_1'(\frac{1}{2})-c_2'(\frac{1}{2})}{\D^3}+\frac{(n-1)}{\D^3}\left(\frac{A'(\frac{1}{2})c_1''(\frac{1}{2})-A''(\frac{1}{2})c_1'(\frac{1}{2})}{A'(\frac{1}{2})^2}\right)+O(\frac{1}{\D^4})\,.
\ee
Notice that the correction begins at $1/\D^2$.
This leads to the conclusion that the $n$-dependence is only through the last term in the above equation and the rest of the pieces are universal. Also observe that $B(z)$ in the block does not enter the final expression.
We find 
\be
\begin{split}
&A'(\frac{1}{2})=2\sqrt{2}\,, \quad c_1'(\frac{1}{2})=\frac{d(6-d)-4}{16\sqrt{2}}\,,\quad c_1''(\frac{1}{2})=\frac{5 ((d-6) d+4)}{16 \sqrt{2}}\,,\quad A''(\frac{1}{2})=-2 \sqrt{2}\,,\\
& c_2'(\frac{1}{2})=\frac{1}{512} \left(d \left(d \left(d \left(-2 \sqrt{2} d+3 d+32 \sqrt{2}-36\right)-136 \sqrt{2}+108\right)+32 \left(4 \sqrt{2}+3\right)\right)-16 \left(2 \sqrt{2}+9\right)\right)\,,.
\end{split}
\ee
This explains why in the numerical plots, we found $g_i$'s approaching $2\sqrt{2}\approx 2.828$. Furthermore, since for non-zero spins, the leading order block is the same (up to some overall $\ell$ dependent factor which drops out in $g_i$), for non-zero spins too we will find the same result at leading order, namely $g_i\approx 2\sqrt{2}$. See Appendix \ref{gi_deri} for derivation in general case. This essentially means:

{\it For large $\Delta$ operators, the structure of the conformal block expansion is that of a cyclic polytope in any dimensions.}

We then explicitly have for the $\ell=0$ case
\be
g_n=2 \sqrt{2}-\frac{d^2-6 d+4}{16 \sqrt{2} \Delta ^2}+\frac{24-d \left(d \left(\sqrt{2} d-6 \sqrt{2}-3\right)+4 \sqrt{2}+30\right)+3((d-6) d+4) (n-1)}{32 \D^3}+O\left(\frac{1}{\D^4}\right)\,.
\ee
This means that the universal part is
\be
g_n^{univ}=2 \sqrt{2}-\frac{d^2-6 d+4}{16 \sqrt{2} \Delta ^2}+\frac{24-d \left(d \left(\sqrt{2} d-6 \sqrt{2}-3\right)+4 \sqrt{2}+30\right)}{32 \D^3}+O\left(\frac{1}{\D^4}\right)\,,
\ee
while the non-universal $n$-dependent part is
\be
g_n^{non-univ}=(n-1)\times\frac{3((d-6) d+4)}{32 \D^3}+O\left(\frac{1}{\D^4}\right)\,.
\ee
To be explicit
\begin{eqnarray}
g_n|_{d=2}&\approx& 2.828+\frac{0.177}{\D^2}+\frac{0.031[-12.686-12(n-1)]}{\D^3}+O(1/\D^4)\,,\\
g_n|_{d=3}&\approx& 2.828+\frac{0.221}{\D^2}+\frac{0.031[-17.787-15(n-1)]}{\D^3}+O(1/\D^4)\,,\\
g_n|_{d=4}&\approx& 2.828+\frac{0.177}{\D^2}+\frac{0.031[-25.373-12(n-1)]}{\D^3}+O(1/\D^4)\,,\\
g_n|_{d=6}&\approx& 2.828-\frac{0.177}{\D^2}+\frac{0.031[-81.941+12(n-1)]}{\D^3}+O(1/\D^4)\,.
\end{eqnarray}
Notice that $c_1'(\frac{1}{2})$ changes sign at $d=6$. Furthermore notice that $c_1'(\frac{1}{2})|_{d=2}=c_1'(\frac{1}{2})|_{d=4}=-c_1'(\frac{1}{2})|_{d=6}=\frac{1}{4\sqrt{2}}.$ For $d\geq 6$, the leading $O(1/\D^2)$ correction is negative. Unless $\D\gg \frac{d-2}{2}$, the $O(1/\D^2)$ and $O(1/\D^3)$ terms are comparable. However, each can still be much smaller than the $2\sqrt{2}$ term. For instance for $d=4,n=1$ both $O(1/\D^2)$ and $O(1/\D^3)$ terms are $<0.1$.

\subsection{Spin blocks}

\begin{figure}[hbt!]
  \centering
 
    \includegraphics[width=0.6\linewidth]{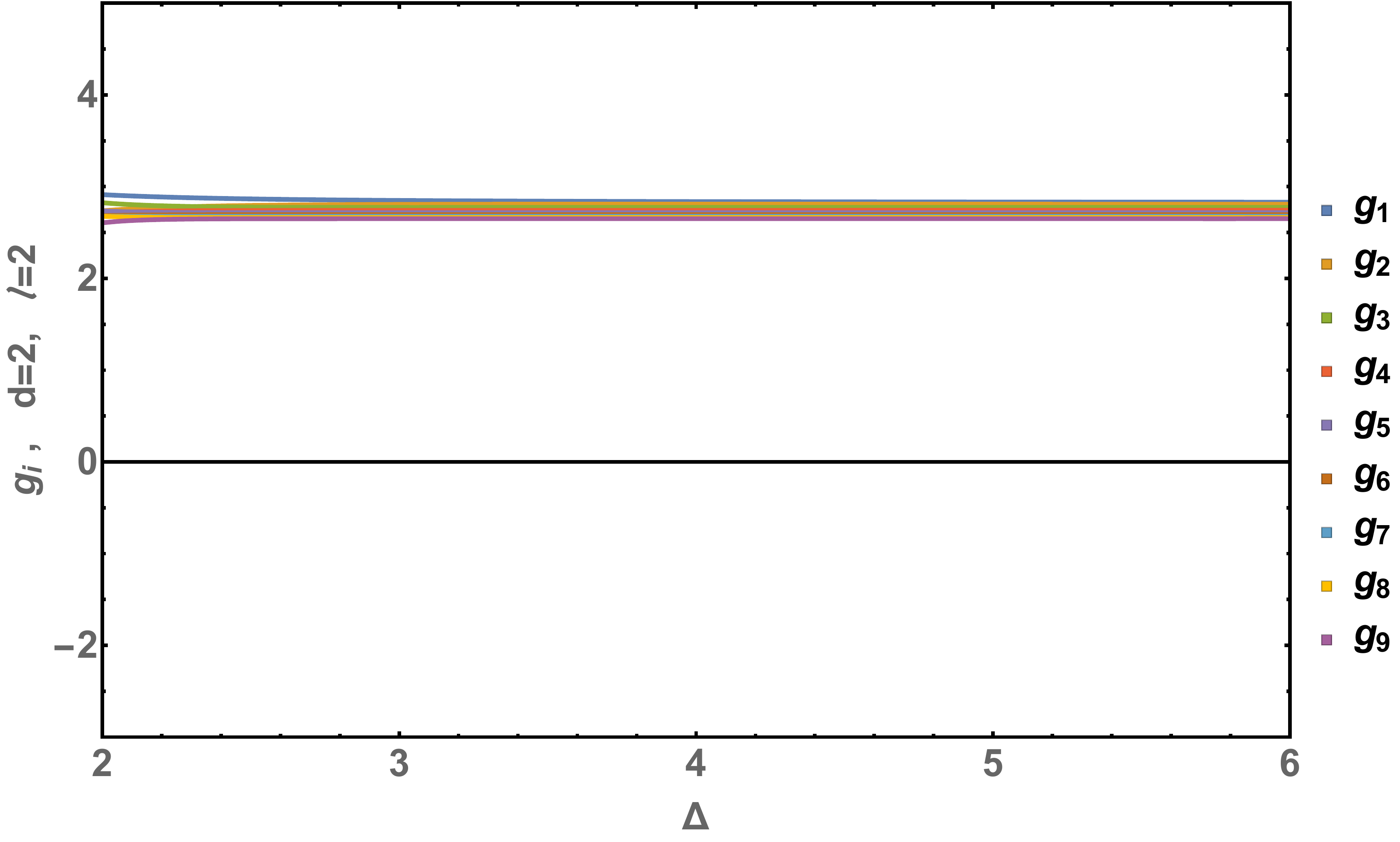}
    \caption{$g_i~vs~\D $ for $d=2,~\ell=2$ }
 
  \label{fig:spin2block}
\end{figure}

Similar analysis can be done for spin blocks for example we include here $\ell=2$ for $d=2$ . The plots look similar  for $d=3,4$ (Figure [\ref{fig:spin2block}]) .

Note that all the $g_i$'s are positive and almost $2\sqrt{2}$ for $\ell=2$. The situation gets better for $\ell=4,6,8$ and so on, since the unitary bounds are $\D\geq d+\ell-2$ and the $\D\gg~ \frac{d-2}{2},\ell$ analysis is valid here.


\subsection{Weighted Minkowski sum}

The Minkowski sum  of set of vectors $X_i,~i=1,2, \dots, n$ ,  is defined as the set of vectors which can be written as the sum of a vector in $X_i$ \cite{bai}
\be
X_1+X_2+\dots +X_n:=\left\{x_1+x_2+\dots +x_n| x_i\in X_i ,~ \forall i\right\} 
\ee

Note that Minkowski sum of polytopes is also a polytopes. Now if $P_i,~i=1,2, \dots, n$ are polytopes in $\mathbb{R}^d$ then their weighted Minkowski sum  $P(\l_i)$ is defined as \cite{bai}
\be
P(\l_i)=\l_1 P_1+\l_2 P_2+\dots +\l_n P_n;~ \l_i\geq 0, ~\forall i
\ee

As we mentioned the condition $g_i>0$ is sufficient but not necessary. When $g_i<0$ we resort to numerically checking $\left\langle \vec{G}_{d,\Delta_1,\ell_1},\vec{G}_{d,\Delta_2,\ell_2},~\dots \vec{G}_{d,\Delta_i,\ell_i}\right\rangle>0$. We have found  for each spin upto some small caveat pertaining to scalars close to the unitarity bound (see appendix (\ref{smallD})) that there is a cyclic polytope structure for unitary theories. 
After adding spin, the resulting polytope is $P(c_\ell)=c_0 P_0+c_2 P_2+c_4 P_4+\dots$,  where $P_\ell$ is the cyclic polytope structure following from the spin-$\ell$ block vectors. So we get the weighted Minkowski sum \cite{bai} of cyclic polytopes which itself is not necessarily a cyclic polytope (as we explicitly demonstrate in the Appendix \ref{somecheck}). However if we work with the approximate blocks, as we have shown above, the result is again a cyclic polytope irrespective of spin sum. 

This is the motivation for looking at the large $\Delta$ limit in order to derive some intuition about what the cyclic polytope picture can teach us about bootstrap.


\section{$N = 1$: Bounds on leading operator} \label{subsec:N=1}

A general constraint arising from $\mathbf{U}\{\D\}\cap \mathbf{X}[\D_\phi]$ is then given by a $2n\times 2n$ matrix whose last $n$ entries are the columns from the basis vectors of the unitary polytope. As a simple example, we can demonstrate some generalizations of the bounds for $N=1$ in (projecting through identity) in \cite[eq. 7.9]{nima} for higher dimensions.
The crossing plane $\mathbf{X}$ is one-dimensional,
\be
\mathbf{X}=\left(\begin{array}{cc}{\mathcal{A}^{0}} & {\mathcal{A}^{2}} \\ {1} & {0} \\ {4 \Delta_{\phi}} & {0} \\ {0} & {0} \\ {\frac{16}{3}\left(\Delta_{\phi}-4 \Delta_{\phi}^{3}\right)} & {4 \Delta_{\phi}}\end{array}\right)\,.
\ee
The polytope is 3-dimensional. The facets can be found by using (\ref{eq:cyclic_face})
\be
(\mathbf{G}_{d,\D_0,0}, \mathbf{G}_{d,\D_i,0},\mathbf{G}_{d,\D_{i+1},0}), \quad(\mathbf{G}_{d,\D_i,0},\mathbf{G}_{d,\D_{i+1},0},\mathbf{G}_{d,\D_{\infty},0})
\ee

Here the identity operator $\mathbf{G}_{d,\D_{0},0}=(1,0,0,0)$ and $\infty$ is  $\mathbf{G}_{d,\D_{\infty},0}=(0,0,0,1)$. No other operator in between $\D_i$ and $\D_{i+1}$ and we order them as $\D_i<\D_{i+1}$. The crossing plane $\mathbf{X}$ intersects with the facets  if and only if equation (\ref{eq:intsec}) is satisfied. As we mentioned earlier  it is sometimes simple and useful to take the projection through some special point \cite{nima}.

\subsection{Projection through identity}
We project our three-dimensional cyclic polytope through the identity $\mathbf{G}_{d,\D_0,\ell_0}=(1,0,0,0)$ . As mentioned in \cite{nima} that the necessary conditions one can get by forcing  $\langle\mathbf{X}, 0, \D \rangle$ to be zero . The crossing plane intersects the unitary cyclic polytope at two points which are given by the two solutions $\D_{-},~\D_{+}$ of 
\be
\langle\mathbf{X}, 0, \D \rangle=0\,.
\ee
This tells us that there must exist at least one operator with dimension $\D$ such that
\be
\D_{-}<\D< \D_{+}\,.
\ee
This also implies that the gap between the identity and the leading operator has to be smaller than $\D_{+}$. Below we tabulate $\D_{+}, \D_{-}$ for various $\Dphi$:
\begin{table}[hbt!]
\centering
\begin{minipage}{0.5\textwidth}
\centering
\begin{tabular}{|c |c| c|}\hline
$\Dphi$ & $\D_{+}$ &  $\D_{-}$\\\hline
 0.1 & 1.29591 & 0.111592 \\\hline
 0.2 & 1.58656 & 0.237821 \\\hline
 0.3 & 1.87438 & 0.371258 \\\hline
 0.4 & 2.16053 & 0.508445 \\\hline
 0.5 & 2.44564 & 0.647694 \\\hline
 0.6 & 2.73006 & 0.788127  \\\hline
\end{tabular}
\caption{List of $\D_{+} , \D_{-}$ for various  $\Dphi$,  $d=1$ }
\label{tab:1d_dpdm}
\end{minipage}%
\begin{minipage}{0.5\textwidth}
\centering
\begin{tabular}{|c |c| c|}\hline
$\Dphi$ & $\D_{+}$ &  $\D_{-}$\\\hline
 0.1 & 1.20812 & 0.120067 \\\hline
0.125 & 1.2826 & 0.151471 \\\hline
 0.2 & 1.50422 & 0.247921 \\\hline
 0.3 & 1.79662 & 0.380248 \\\hline
 0.4 & 2.08665 & 0.515341 \\\hline
 0.5 & 2.37507 & 0.652232 \\\hline
\end{tabular}
\caption{List of $\D_{+} , \D_{-}$ for various  $\Dphi$,  $d=2$ }
\end{minipage}
\end{table}

\begin{table}[hbt!]
\centering
\begin{minipage}{0.5\textwidth}
\centering
\begin{tabular}{|c |c| c|}\hline
$\Dphi$ & $\D_{+}$ &  $\D_{-}$\\\hline
 0.5 & 2.26058 & 0.697759 \\\hline
 0.518 & 2.31459 & 0.715473  \\\hline
 0.6 & 2.55882 & 0.807958 \\\hline
 0.7 & 2.85357 & 0.934263 \\\hline
 0.8 & 3.14593 & 1.06673 \\\hline
 0.9 & 3.43657 & 1.20214 \\\hline
 1. & 3.72592 & 1.33921 \\\hline
\end{tabular}
\caption{List of $\D_{+} , \D_{-}$ for various  $\Dphi$,  $d=3$ }
\end{minipage}%
\begin{minipage}{0.5\textwidth}
\centering
\begin{tabular}{|c |c| c|}\hline
$\Dphi$ & $\D_{+}$ &  $\D_{-}$\\\hline
 1. & 3.61311 & 1.3043 \\\hline
 1.2 & 4.20493 & 1.58012 \\\hline
 1.4 & 4.789 & 1.85969 \\\hline
 1.6 & 5.36839 & 2.14053 \\\hline
 1.8 & 5.94472 & 2.42204 \\\hline
 1.9 & 6.23204 & 2.56297   \\\hline
 2. & 6.51892 & 2.70398 \\\hline
\end{tabular}
\caption{List of $\D_{+} , \D_{-}$ for various  $\Dphi$,  $d=4$ }
\label{tab:4d_dpdm}
\end{minipage}
\end{table}
It is possible to further constrain the $\D_+$ and $\D_-$ including spin analysis from the technique of $N=2$ analysis given in section (\ref{sec:N=2}). We will study this in detail in a forthcoming publication \cite{YSWZ}. In appendix (\ref{ap:kink_spin}) we have made a preliminary study of the effects of adding spin to the analysis of bound on the dimension of the leading scalar.

\subsection{ $\D_{+}$ and $\D_{-}$ for large $\Dphi$}
Using eq (\ref{block}) with the approximations of the ${}_3F_2$ given in section 2.2,  \footnote{ For $O(\frac{1}{\Dphi^n})$ one has to add $O(\frac{1}{\D^n})$ corrections to eq (\ref{blockapprox}) using \cite{luke1}.}  we can find analytic solutions for $\D_{+}$ and $\D_{-}$  for large $\Dphi$:
\be \label{dpdm}
\begin{split}
\D_{+}=&~2 \sqrt{2} \Delta _{\phi }+\frac{\left(2 \sqrt{2}-3\right) d+6}{4 \sqrt{2}}+\frac{12-d (d+6)}{128 \sqrt{2} \Delta _{\phi }}-\frac{3 (d (d (d+2)-44)+88)}{2048 \sqrt{2} \Delta _{\phi }^2}\\
&+\frac{d (-d (d+6) (37 d-282)-7392)+15216}{131072 \sqrt{2} \Delta _{\phi }^3}+O\left(\frac{1}{\Delta _{\phi }^4}\right)\\
\D_{-}=&~\sqrt{2} \Delta _{\phi }+\frac{1}{8} \left(4-3 \sqrt{2}\right) d-\frac{(d-6) d+12}{64 \sqrt{2} \Delta _{\phi }}-\frac{3 \left((d-4)^2 d-32\right)}{512 \sqrt{2} \Delta _{\phi }^2}\\
&+\frac{d (d ((372-37 d) d-1188)+480)-1680}{16384 \sqrt{2} \Delta _{\phi }^3}+O\left(\frac{1}{\Delta _{\phi }^4}\right)
\end{split}
\ee
This formula is even very good approximation for $\Dphi>1$. For comparison we give a plot for $\D_{+}$ and $\D_{-}$ vs $\Dphi$ calculated using exact block and eq (\ref{dpdm}).\\
We can also expand around  $\D_{\Phi}=\Dphi+a$

\be \label{dpdm2}
\begin{split}
\D_{+}=&~2 \sqrt{2} \Delta_\Phi +\frac{1}{8} \left(-16 \sqrt{2} a+\left(4-3 \sqrt{2}\right) d+6 \sqrt{2}\right)+\frac{12-d (d+6)}{128 \sqrt{2} \Delta_\Phi }\\
&+\frac{-16 a (d (d+6)-12)-3 (d (d (d+2)-44)+88)}{2048 \sqrt{2} \Delta_\Phi ^2}\\
&+\frac{-1024 a^2 (d (d+6)-12)-384 a (d (d (d+2)-44)+88)+d (-d (d+6) (37 d-282)-7392)+15216}{131072 \sqrt{2} \Delta_\Phi ^3}\\
&+O\left(\frac{1}{\Delta _{\Phi }^4}\right)\\
\D_{-}=&~\sqrt{2} \Delta _{\Phi }+\left(\frac{1}{8} \left(4-3 \sqrt{2}\right) d-\sqrt{2} a\right)-\frac{(d-6) d+12}{64 \sqrt{2} \Delta _{\Phi }}+\frac{-8 a ((d-6) d+12)-3 \left((d-4)^2 d-32\right)}{512 \sqrt{2} \Delta _{\Phi }^2}\\
&+\frac{-256 a^2 ((d-6) d+12)-192 a \left((d-4)^2 d-32\right)+d (d ((372-37 d) d-1188)+480)-1680}{16384 \sqrt{2} \Delta _{\Phi }^3}\\
&+O\left(\frac{1}{\Delta _{\Phi }^4}\right)
\end{split}
\ee

\begin{figure}[hbt!]
  \centering
  \begin{subfigure}{0.7\linewidth}
    \includegraphics[width=\linewidth]{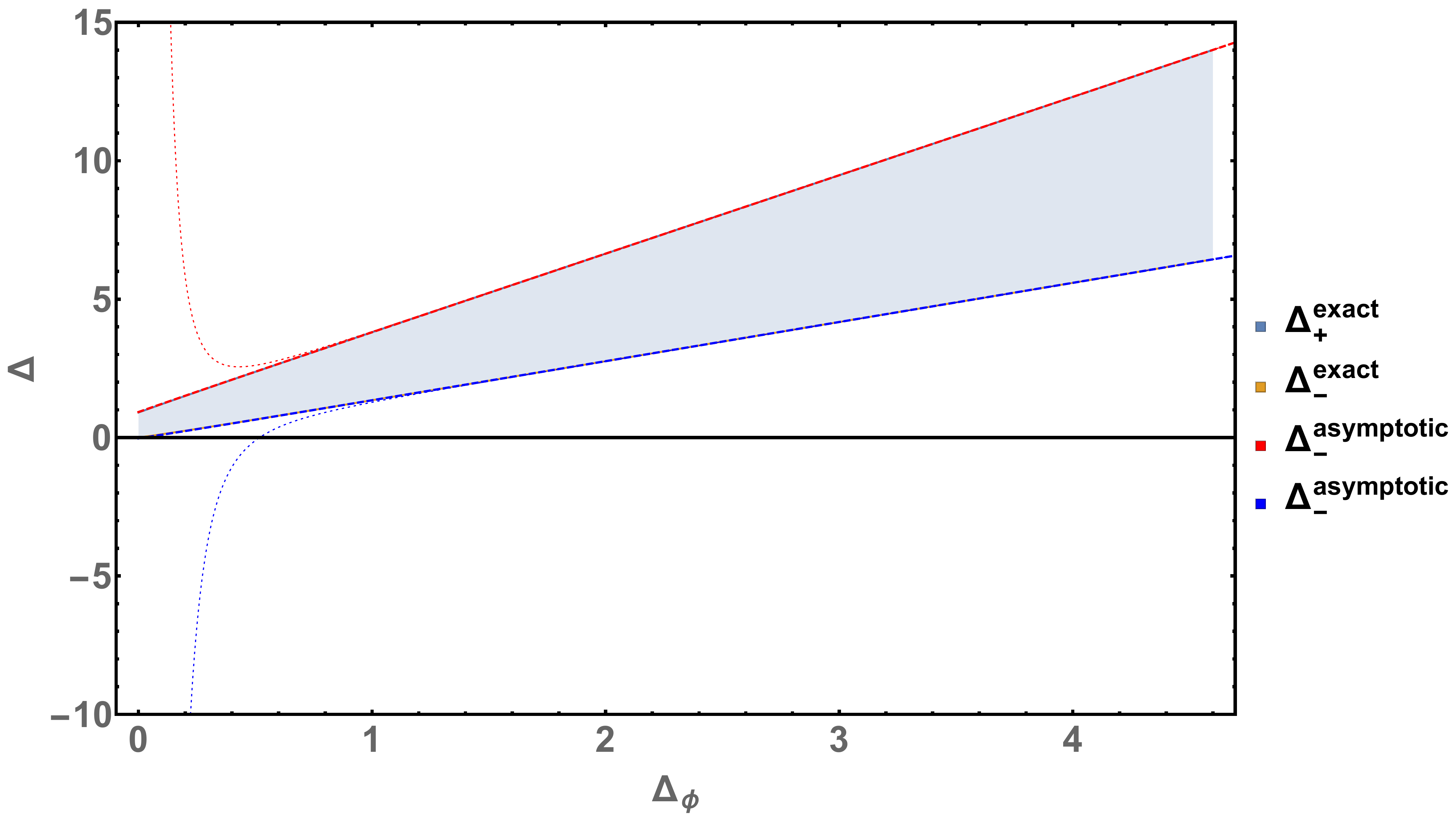}
    \caption{$\D_{+} , \D_{-}$ vs $\Dphi$ for $d=2$ }
  \end{subfigure}
  \begin{subfigure}{0.7\linewidth}
    \includegraphics[width=\linewidth]{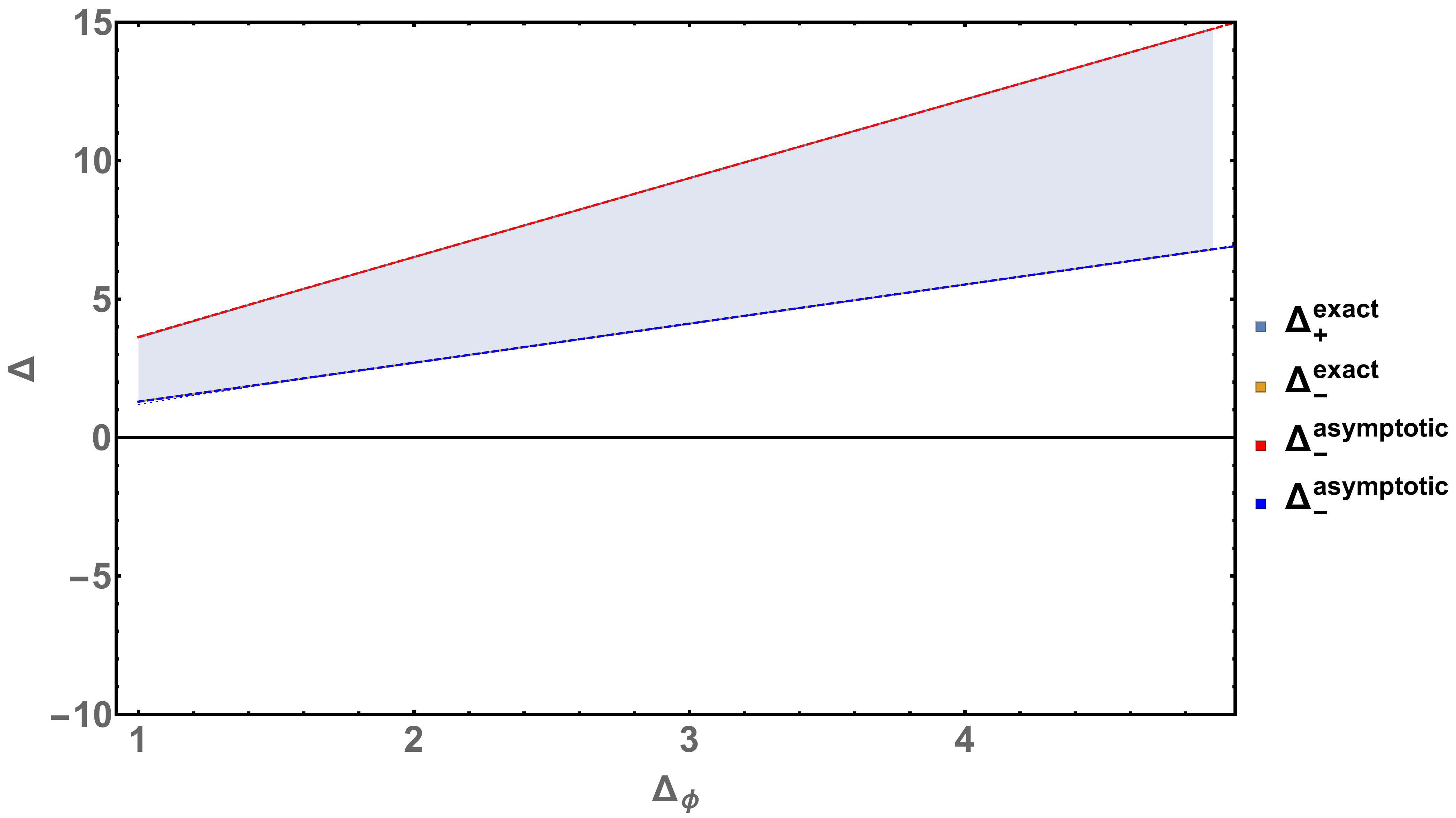}
    \caption{$\D_{+} , \D_{-}$ vs $\Dphi$ for $d=4$ }
  \end{subfigure}
  \caption{Solid lines represent $\D_{+} , \D_{-}$ calculated using exact block while dashed lines are eq (\ref{dpdm2}),dashed lines represent $a=1$ and dotted are $a=0$. We have included here only $d=2,4$ but the feature is similar for $d=1,3$.}
  \label{fig:DpDm_withasym}
\end{figure}

One can easily see from Figure (\ref{fig:DpDm_withasym}) that  eq (\ref{dpdm2}) with $a=0$ i.e \ref{dpdm} is very good approximation for $\Dphi>1$, while it deviates a lot from the exact numbers for $\Dphi<1$. Also \ref{dpdm2} is very good approximation even for $\Dphi<1$ for the choice of $a=1$. One can optimize the value of $a$ for better numerical agreement for $\Dphi<1$, which we have not done here.

\subsection{Interpretation of $\D_{+}$ and $\D_{-}$ from numerical bootstrap}

From crossing symmetry equation (\ref{cross}, \ref{cross_diag}) in diagonal limit we have, 
\be
\begin{split}
\sum_{\D,\ell}C_{\D,\ell} \mathcal{F}_{d,\D,\ell}(z)=1\,,
\end{split}
\ee
where $\mathcal{F}_{d,\D,\ell}(z)$ is given in equation (\ref{daigcross}). 
We can write
\be\label{F=0}
\begin{split}
\sum_{\D,\ell}C_{\D,\ell} \pd^2_z \mathcal{F}_{d,\D,\ell}(z)|_{z=1/2}=0\,,
\end{split}
\ee
From figure (\ref{fig:llp_DpDm}), it is clear that $\pd^2_z \mathcal{F}_{d,\D,0}(z)|_{z=1/2}$ changes its sign at indicated values. At least there should be one operator below $\D_{+}$ (the larger value) in order to satisfy equation (\ref{F=0}).

\begin{figure}[h!]
  \centering
  \begin{subfigure}[b]{0.45\linewidth}
    \includegraphics[width=\linewidth]{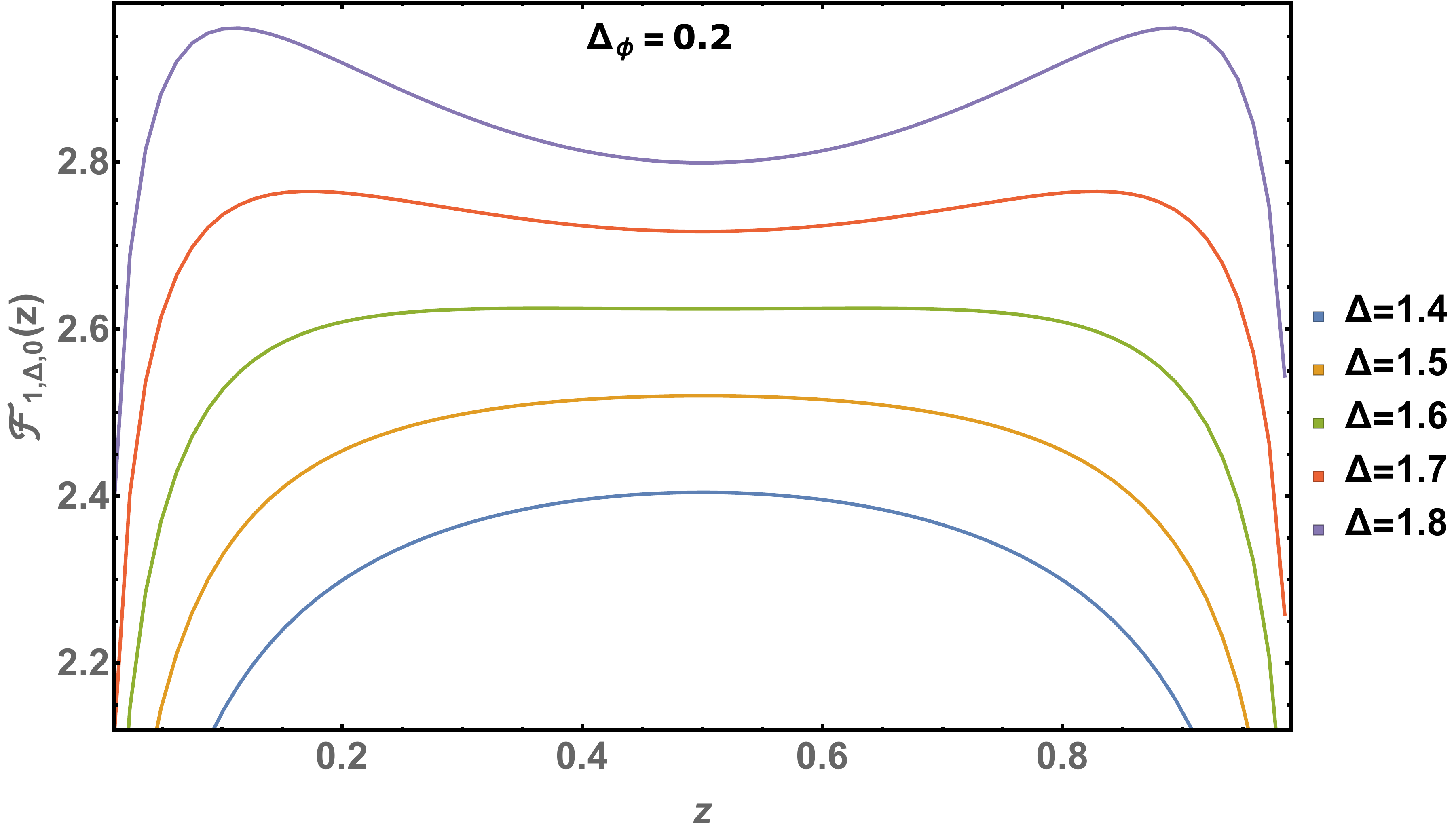}
    \caption{$\mathcal{F}_{d,\D,0}(z)$ vs $z$ for $d=1$ }
  \end{subfigure}
  \begin{subfigure}[b]{0.45\linewidth}
    \includegraphics[width=\linewidth]{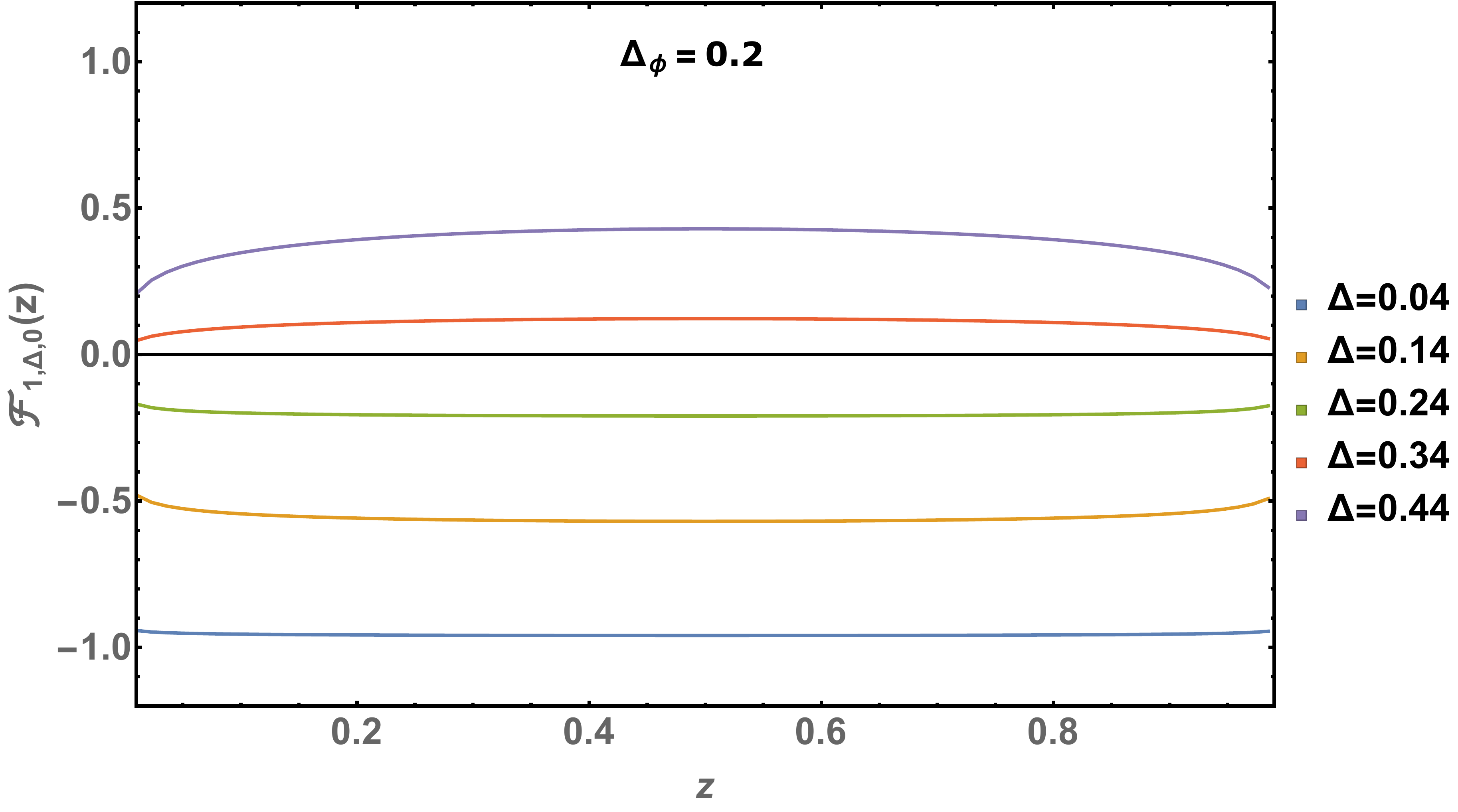}
    \caption{$\mathcal{F}_{d,\D,0}(z)$ vs $z$ for $d=1$}
  \end{subfigure}
  \begin{subfigure}[b]{0.45\linewidth}
    \includegraphics[width=\linewidth]{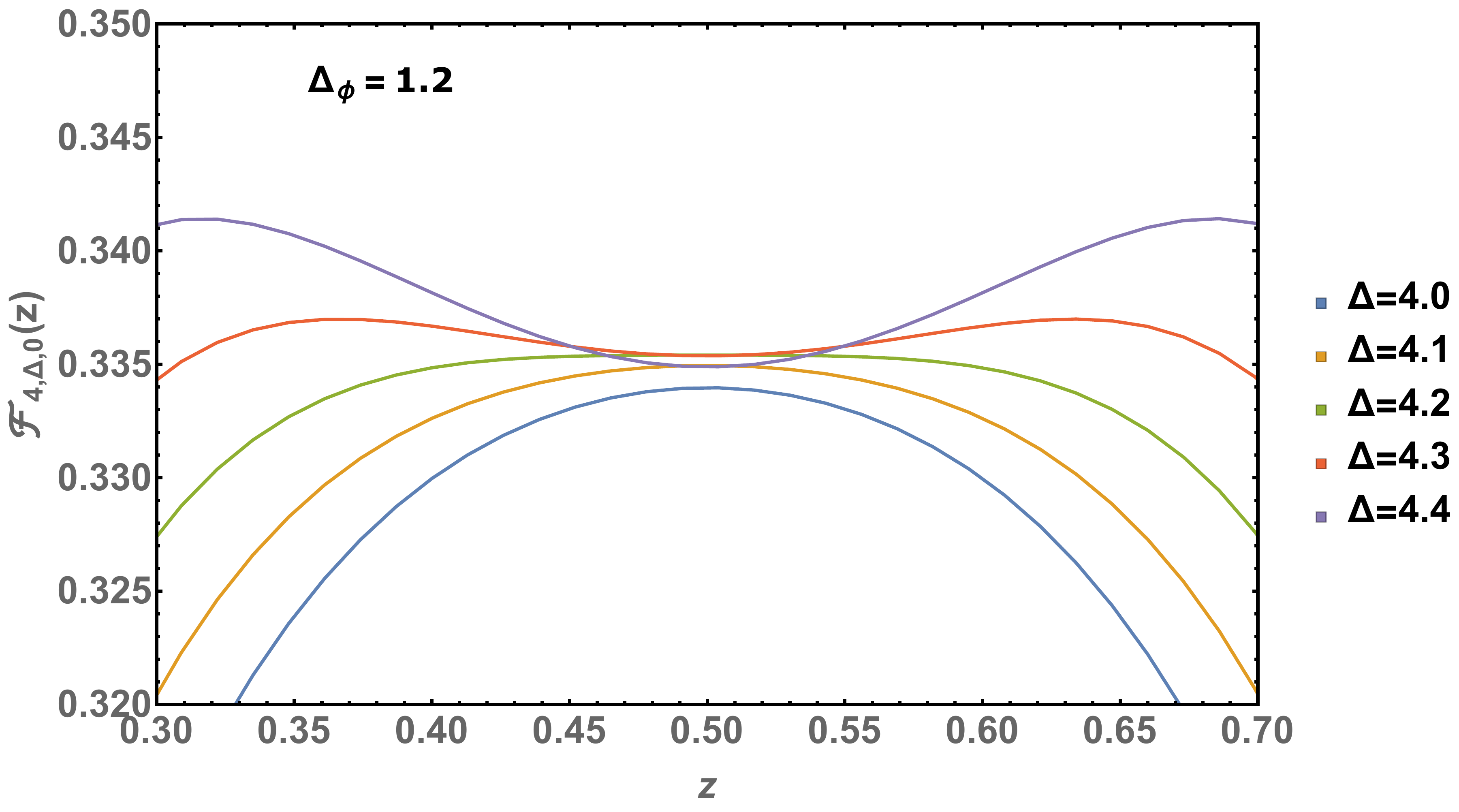}
    \caption{$\mathcal{F}_{d,\D,0}(z)$ vs $z$ for $d=4$}
  \end{subfigure}
  \begin{subfigure}[b]{0.45\linewidth}
    \includegraphics[width=\linewidth]{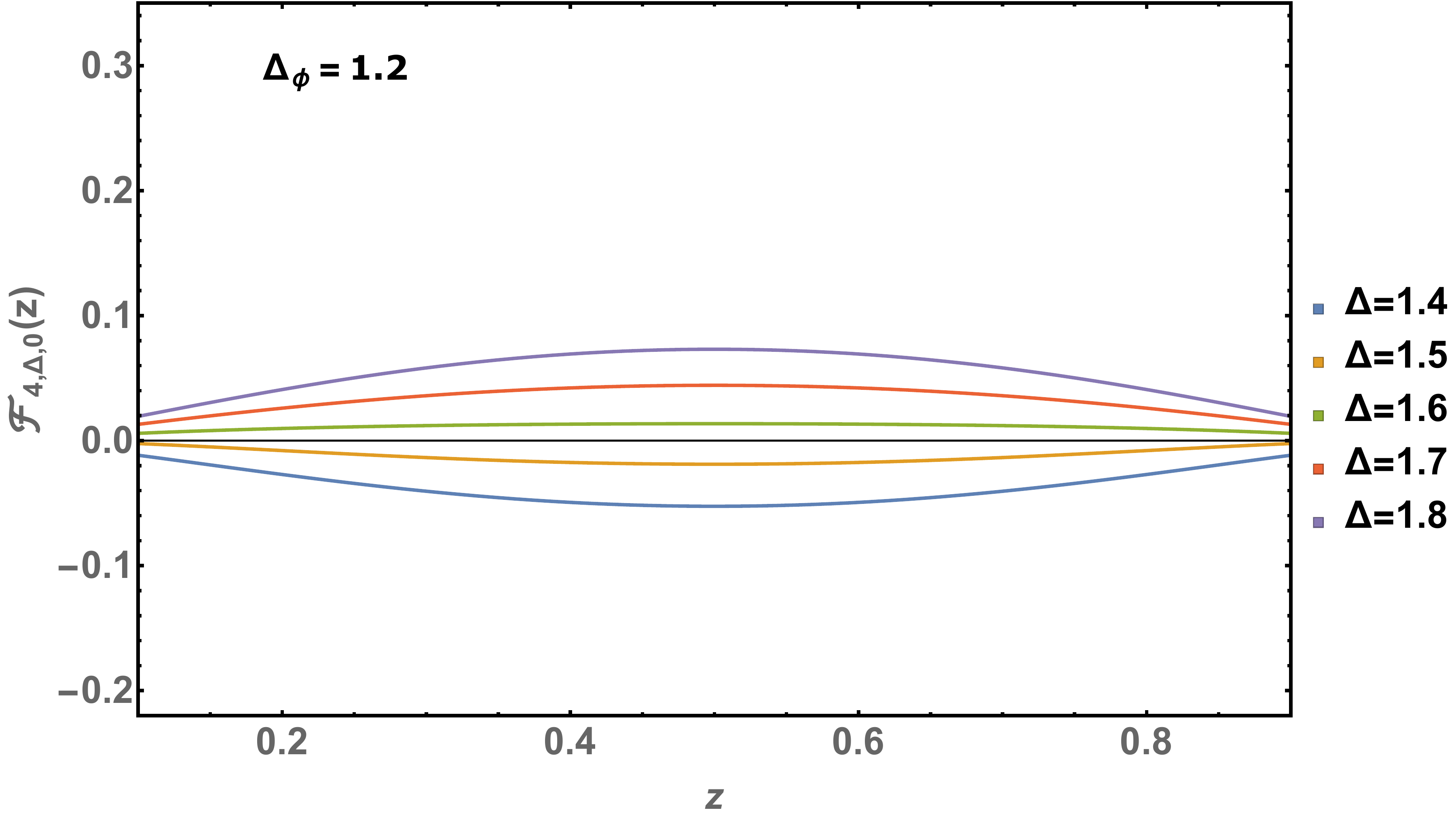}
    \caption{$\mathcal{F}_{d,\D,\ell}(z)$ vs $z$ for $d=4$ }
  \end{subfigure}
  \caption{$\mathcal{F}_{d,\D,0}(z)$ vs $z$ to show $\D_{+}$ and $\D_{-}$ from numerical bootstrap}
  \label{fig:llp}
\end{figure}

\begin{figure}[hbt!]
  \centering
  \begin{subfigure}[b]{0.45\linewidth}
    \includegraphics[width=\linewidth]{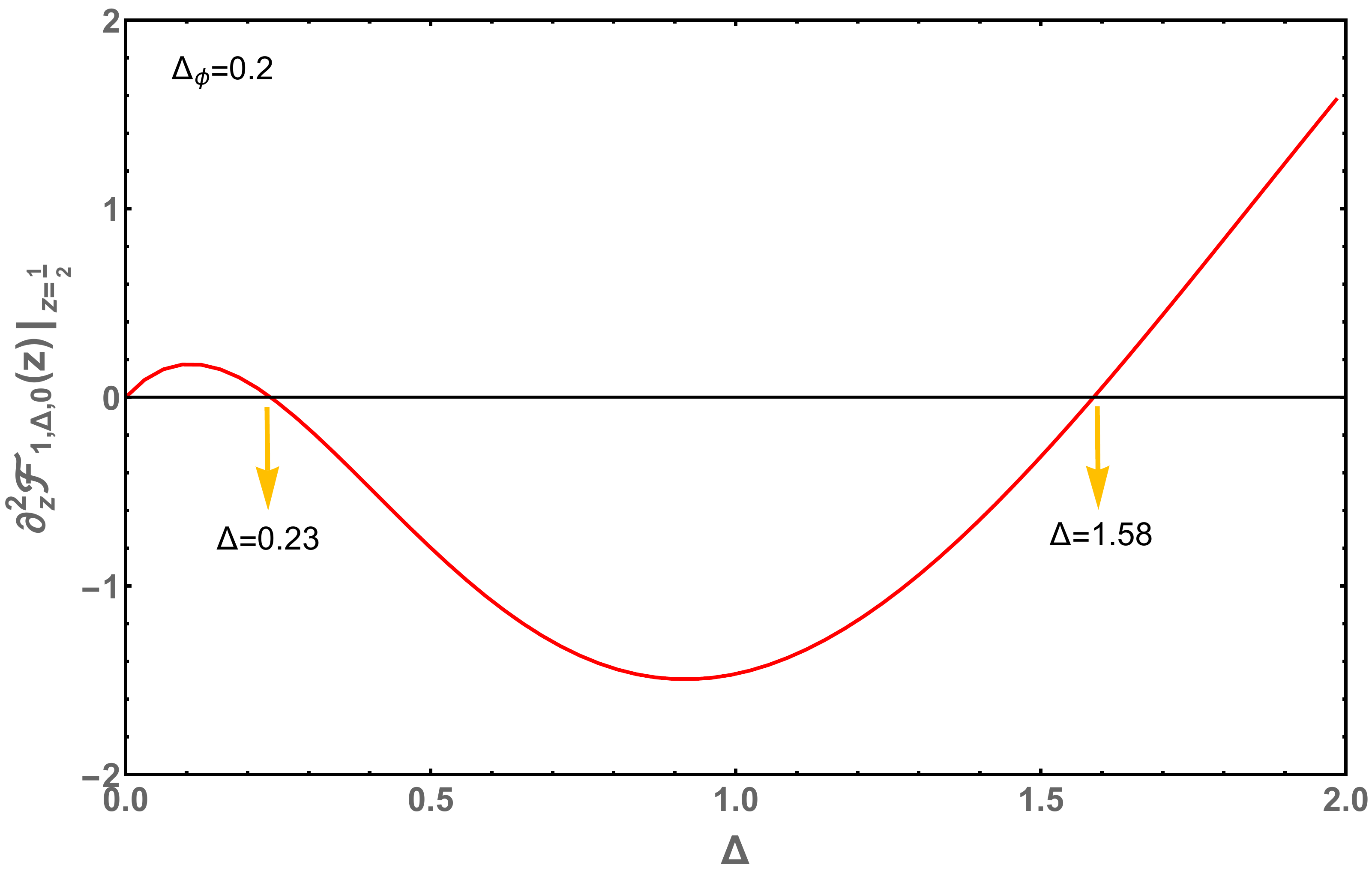}
    \caption{$\partial^2_z\mathcal{F}_{d,\D,0}(z)|_{z=1/2}$ vs $\D$ for $d=1$ }
  \end{subfigure}
  \begin{subfigure}[b]{0.45\linewidth}
    \includegraphics[width=\linewidth]{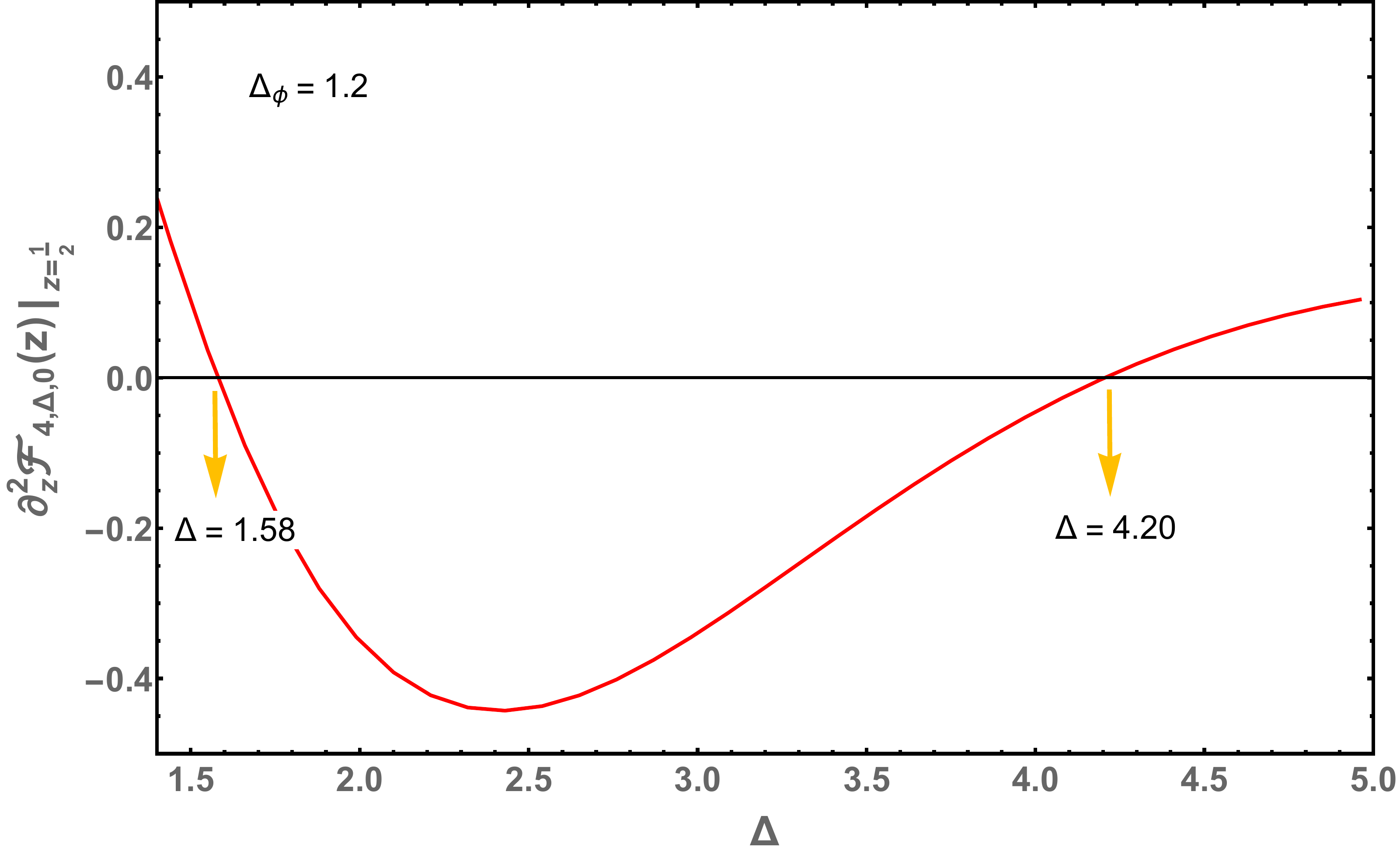}
    \caption{$\partial^2_z\mathcal{F}_{d,\D,0}(z)|_{z=1/2}$ vs $\D$ for $d=4$}
   \end{subfigure}
  \caption{$\partial_z^2\mathcal{F}_{d,\D,0}(z)|_{z=1/2}$ vs $\D$. See table \ref{tab:1d_dpdm} and table \ref{tab:4d_dpdm} for values of $\D_{+}$ and $\D_{-}$. The indicated values in the plot is exactly matched with the value given in the tables}
  \label{fig:llp_DpDm}
\end{figure}

A peculiar feature that is captured by both the analysis we are doing in this section as well as the polytope method, is that there is a third root of the $N=1$ determinant (or equivalently $\partial^2_z {\mathcal F}_{d,\D,0}(z)|_{z=\frac{1}{2}}$) which allows for a small region $\frac{d-2}{2}<\D<\D_*$ close to the unitarity bound as an allowed region. However, it seems to us that this is somewhat artificial and we will not comment on this further. The main point is that there could always be operators below $\D_-$ as well, but there must be at least one operator (which need not be the leading scalar operator in the OPE which happens if there are operators\footnote{A simple example is the case where in a free theory in $d=4$ we have the external operator to be $\phi^2$ so that the leading scalar operator is $\phi^2$ itself. From table (\ref{tab:4d_dpdm}), it is clear that since $\Delta_{\phi^2}=2$ we have $\Delta_{\phi^2}<\D_-$. } $\D<\D_-$) such that $\D_-<\D<\D_+$ for there to be any chance for a solution.

\section{$N = 2$: Constraints on the first two primary operators}\label{sec:N=2}

\begin{figure}[ht]
  \centering
  \begin{subfigure}{0.4\linewidth}
    \includegraphics[width=0.95\linewidth]{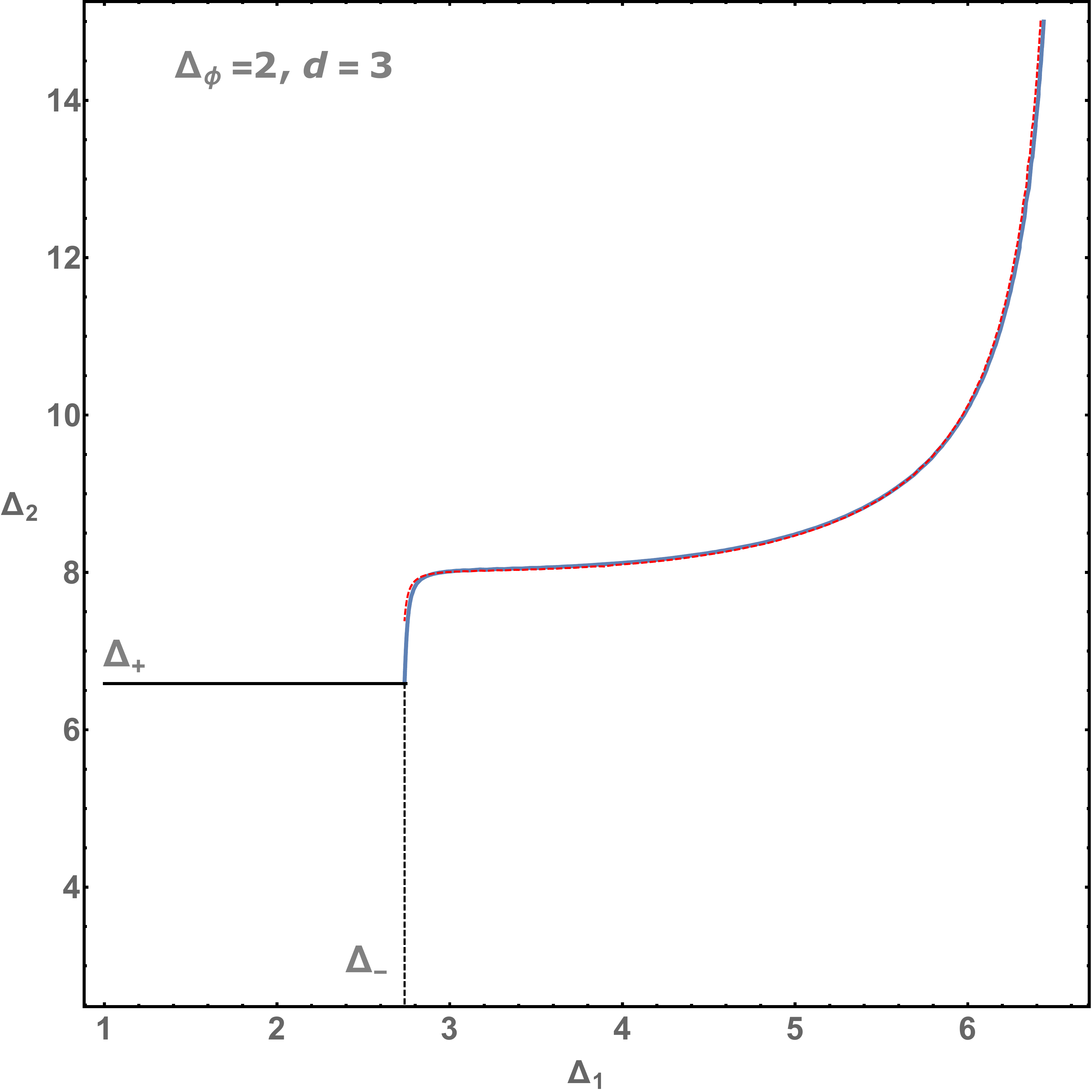}
    
  \end{subfigure}
  \begin{subfigure}{0.4\linewidth}
    \includegraphics[width=0.95\linewidth]{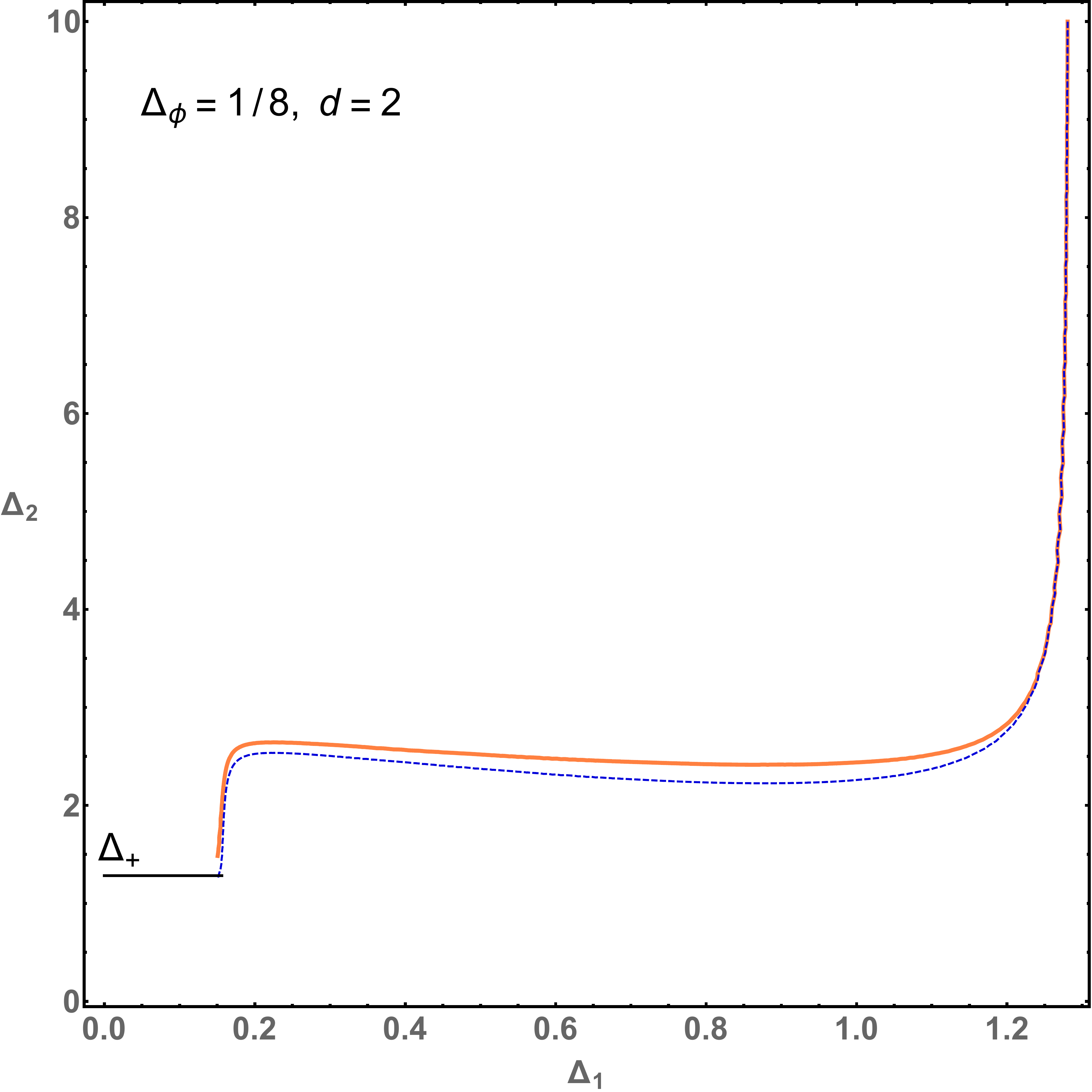}

  \end{subfigure}
  \caption{ (a) Blue line is using exact scalar blocks and the red dashed line is using asymptotic scalar blocks. (b) Orange line is using spin-2 block for $\D_2$ and blue dashed line is using scalar block for $\D_2$.}
    \label{fig:D1D2_2d_ising_model}
\end{figure}

In this section, we consider $N=2$ \textit{i.e.,} the $5$-dimensional polytope. The crossing plane is given in equation (\ref{N=2,X}) and the boundaries can be found using \ref{eq:cyclic_face}.
As before, for simplicity and usefulness we will view the geometry projected through identity \cite{nima}. The constraints for the first  two operators $\D_1,\D_2$ is as follows. For $\D_1<\D_-$ we should have $\D_2<\D_+$, otherwise there will be no operators between ($\D_-,\D_+$) as we observed in the $N=1$ case. And also $\D_1>\D_+$ is not allowed if $\D_1$ is the leading operator. As explained in \cite{nima}, the necessary conditions one would obtain from the observation that  for $\D_- <\D_1<\D_+$, $\D_2$ must be below the curve $\left\langle\mathbf{X},0,\D_1,\D_2\right\rangle=0$ (see figure (\ref{fig:D1D2_2d_ising_model})), otherwise  \ref{eq:intsec} is not satisfied. Explicitly
\be
\begin{split}
\left\langle\mathbf{X},0,\D_1,\D_2\right\rangle=\text{det}\left(
\begin{array}{cccccc}
 1 & 0 & 0 & 1 & 1 & 1 \\
 4 \Delta \phi  & 0 & 0 & 0 & G^{(1)}_{d,\Delta _1,0} & G^{(1)}_{d,\Delta _2,0} \\
 0 & 1 & 0 & 0 & G^{(2)}_{d,\Delta _1,0} & G^{(2)}_{d,\Delta _2,0} \\
 \frac{16}{3} \left(\Delta \phi -4 \Delta \phi ^3\right) & 4 \Delta \phi  & 0 & 0 & G^{(3)}_{d,\Delta _1,0} & G^{(3)}_{d,\Delta _2,0} \\
 0 & 0 & 1 & 0 & G^{(4)}_{d,\Delta _1,0} & G^{(4)}_{d,\Delta _2,0} \\
 \frac{64}{15} \Delta \phi  \left(32 \Delta \phi ^4-20 \Delta \phi ^2+3\right) & \frac{4}{15} \Delta \phi  \left(20-80 \Delta \phi ^2\right) & 4 \Delta \phi  & 0 & G^{(5)}_{d,\Delta _1,0} & G^{(5)}_{d,\Delta _2,0} \\
\end{array}
\right)\,.
\end{split}
\ee

We show in fig. 9 how the curve $\left\langle\mathbf{X},0,\D_1,\D_2\right\rangle=0$ changes if we put the second operator to be one with spin (there is an assumed ordering and in this instance we are assuming that the second operator is one that carries spin, assuming that the cyclic polytope structure is preserved). In figure (\ref{fig:D1D2_2d_ising_model}.a) we have taken $\Dphi=2,~ d=3$ , both $\D_2$ and $\D_1$ blocks are taken to be scalar block for this case. In figure  (\ref{fig:D1D2_2d_ising_model}.b), we have taken the 2d ising model $\Dphi=\frac{1}{8}$ and used the spin $\ell=2$ block for the operator $\D_2$ \textit{i.e} we used $\mathbf{G}_{2,\D_2,2}$ instead of $\mathbf{G}_{2,\D_2,0}$. One can see the plot is only slightly different if one uses scalar block \text{i.e} $\mathbf{G}_{2,\D_2,0}$  for the operator $\D_2$. Had the $\D_2$ operator been a scalar (say by demand) in this case, we would be forced to conclude that $\D_2=4$ is not allowed thereby leading to a contradiction with the known result. Thankfully, from the plot, the spin-2 bound allows for $\D=2$ which is the stress tensor and should be treated as the second operator.


\section{Kink from positive geometry}
In this section we will explain how the red line in fig.1 was obtained. We consider 10 scalar operators $\D_i$ where $\D_0$ is the identity operator and $\D_9$ is the infinity vector. $\D_1=\D_S$ and $\D_i, i\geq 2$ are chosen randomly to be above\footnote{For definiteness $\D_2=\D_S+0.02,\D_3=\D_S+0.12,\D_4=\D_S+0.22,\D_5=\D_S+0.32,\D_6=\D_
S+1.42,\D_7=\D_S+1.52,\D_8=\D_S+2.62$ Changing these spacings do not affect the qualitative features of the plot.} $\D_S$. The intersection conditions eq.\eqref{eq:intsec} are now checked. For $N=1$ we find essentially the same results as obtained in sec. 5.1. However for $N=2$ the result is different. We find that there is a kink type feature in the plot as in fig.1. In fig.1 below the red line, there is always at least some region where the intersection conditions are satisfied (but not above the line). So why does a kink arise?

For concreteness, let us examine the signs of the 
determinants arising from the intersection of the crossing plane with $(\mathbf{v}_0,\mathbf{v}_1,\mathbf{v}_8,\mathbf{v}_9)$. While $det_1=\langle \mathbf{X}, \mathbf{v}_1, \mathbf{v}_8, \mathbf{v}_9\rangle$, $det_2=\langle \mathbf{v}_0, \mathbf{X}, \mathbf{v}_8, \mathbf{v}_9\rangle$ are negative in the range displayed above, $det_3=\langle \mathbf{v}_0, \mathbf{v}_1, \mathbf{X}, \mathbf{v}_9\rangle$ and $det_4=\langle \mathbf{v}_0, \mathbf{v}_1, \mathbf{v}_8, \mathbf{X}\rangle$ are not always negative. The plots in fig.11 are for $det_3, det_4$. Above $\Dphi\approx 0.08$,there is a small region below $\D_S\approx 1.178$ where both are negative but below $\Dphi\approx 0.08$, the region where they are both negative jumps further left giving rise to the kink in fig.1. 

In other dimensions, there are also kinks. For $d=1$, the kink appears $\D\sim 0$, for $d=3$, the kink appears around $\D\sim 0.26$ and for $d=4$, it appears near $\D\sim 0.5$ . The latter two are for non-unitary operators and hence we will not consider them further. When we put in operators with spin, while the qualitative features not affected, the exact location of the kink moves around slightly. However, we have not investigated this in detail as the faces in this case are those of the weighted Minkowski sum which is not that of a cyclic polytope necessarily. If we repeated this analysis with the asymptotic blocks, the kink is no longer as sharp as in fig.1. We leave the search for an analytic (asymptotic) equation for the kink via this approach for future work. 
This is not impossible: a strategy would be to examine carefully the location of the largest zeroes in $det_3,det_4$ by plugging in the asymptotic blocks and studying analytically when they cross each other. We have looked at this problem and our preliminary analysis can be found in appendix \ref{kink}. Another question that we leave for the future is exploring higher $N$ constraints to see what effect it has on the kink.
\begin{figure}[ht]\label{kinkexp}
\begin{tabular}{ccc}
\includegraphics[width=0.3\linewidth]{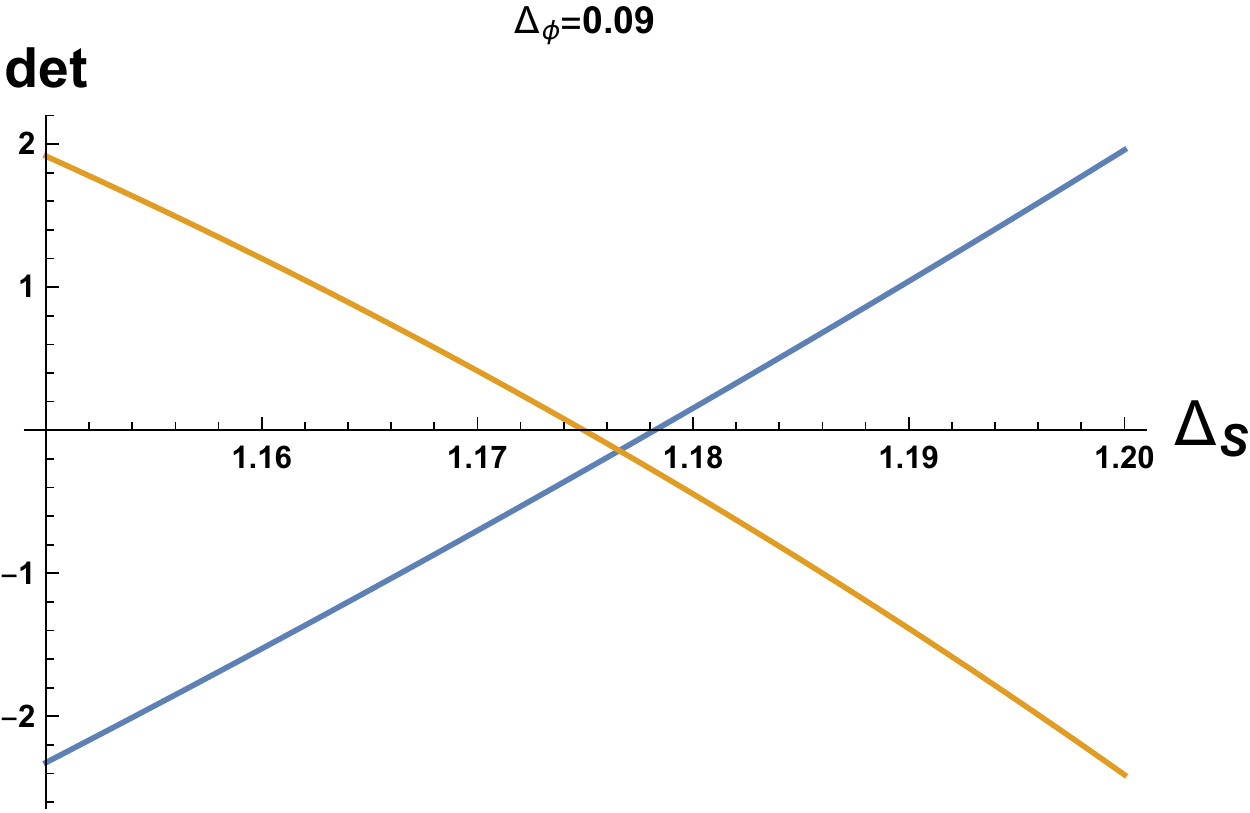}&
\includegraphics[width=0.3\linewidth]{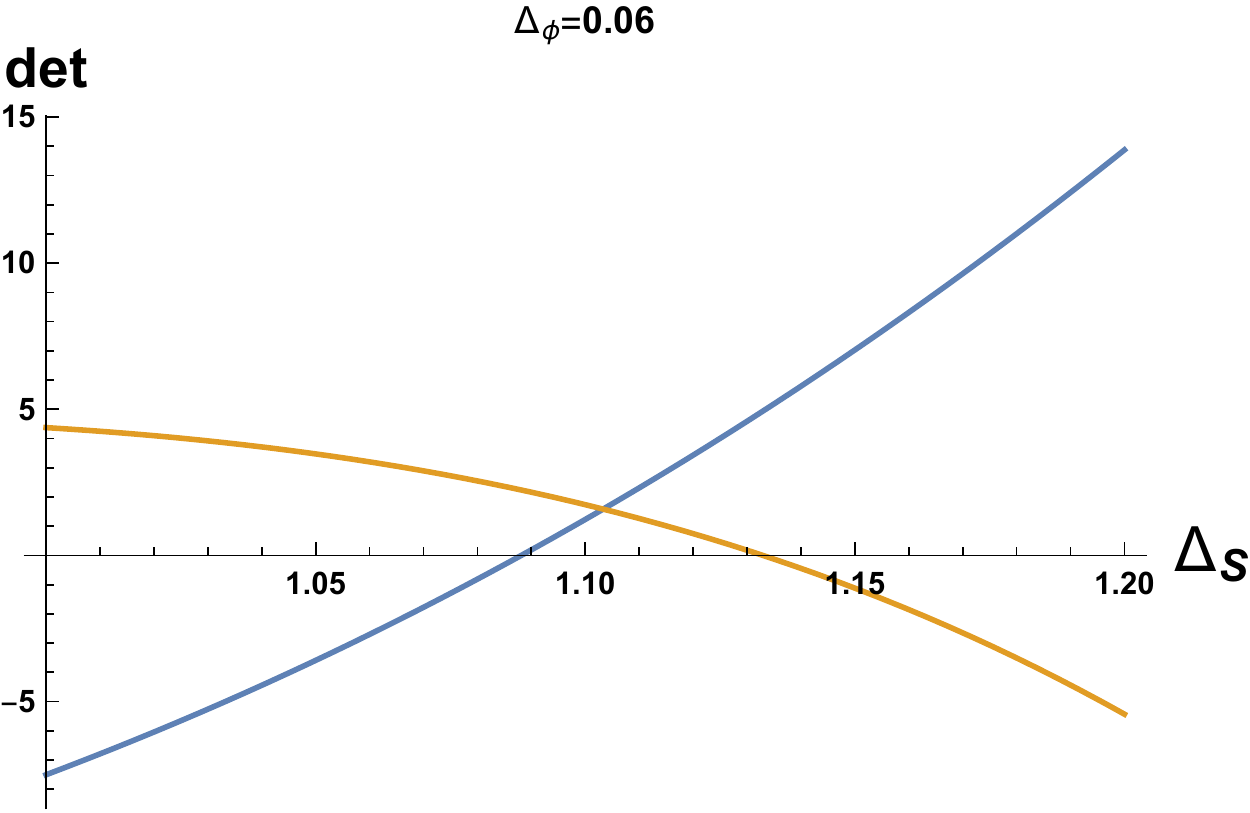} & 
\includegraphics[width=0.3\linewidth]{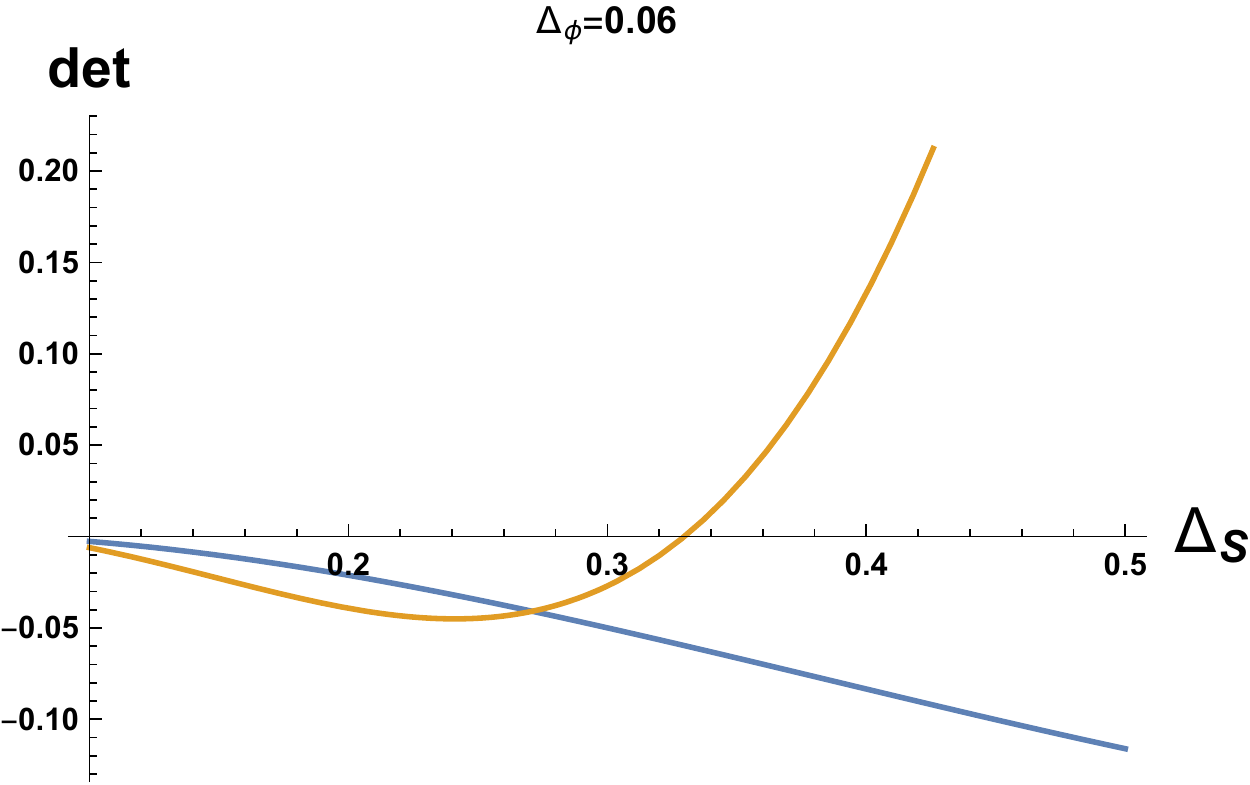}\\
(a)&(b)&(c)
\end{tabular}
\caption{Plots of $det_3, det_4$ as a function of $\D_S$. The blue plot is for $det_3$ while the orange plot is for $det_4$.}
\end{figure}


\section{Discussion}
In this paper we have extended the work of \cite{nima} in several directions. First, we have considered the diagonal limit in arbitrary dimensions and have shown that in the large $\D$ limit, even in the presence of spin, the geometric structure of the expansion in terms of the diagonal blocks of \cite{diag} is that of a cyclic polytope. We presented various analytic results and also initiated a comparison between the usual numerical bootstrap of \cite{rrtv} and the positive geometry constraints.
We conclude with some possible future directions:
\begin{itemize}
\item One can consider more systematically the higher $N$ generalizations of the analysis presented in this paper, especially with the kink picture, also including the contributions from higher spins. This can be done irrespective of the polytope being a cyclic one. It will also be interesting to see if there is some systematic classification of the facets of the weighted Minkowski sum that we found in this paper. More ambitiously, one can also try to develop a systematic asymptotic (in $1/\D$) analytic equation for the kink--in our context we have made a preliminary attempt in appendix C.
\item There is still work to be done in order to connect this geometric picture to the usual numerical bootstrap. The main direction to proceed from here is the generalization of the geometric approach to the non-diagonal case. The usual numerical bootstrap takes the input of both the dimension and spin for the generic off-diagonal blocks ({\it i.e} the block is a function of both $z$ and $\bar{z}$). In that sense the bound obtained from numerical bootstrap is  more constraining. As already pointed out in \cite{nima}, for the off-diagonal case, the simple picture of the polytopes does not apply and it generalizes to tensor products of the simple polytope picture we have handled so far. It would be interesting to see what the generalization has in store for us and how much can still be done analytically using the tensor product of (cyclic) polytopes. 
\item We can also consider the alternate formulation of the bootstrap in \cite{pap1,pap2,pap3,pap4,pap5, gs} using the {\it Polyakov blocks} which are manifestly crossing symmetric and ask about the {\it OPE consistency} in terms of the geometric picture. In that case the consistency conditions arise by demanding the absence of certain spurious poles. These conditions in the diagonal limit schematically read $$\sum_{\Delta,\ell} C_{\Delta,\ell} q_{\Dphi,\Delta,\ell}(s)=0, \quad s=\Dphi+n$$ where $s$ is Mellin variable and $n$ is an integer. Thus they are similar to the $m>0$ conditions listed in eq.(3.3). This would be the natural starting point for a linear programming based approach using Polyakov blocks. It remains to be seen if Polyakov block expansion and cyclic polytopes are in any sense related.
\end{itemize}

\section*{Acknowledgments} We thank Kausik Ghosh, Yu-tin Huang and  Slava Rychkov for discussions and correspondence. A.S. acknowledges support from a DST Swarnajayanti Fellowship Award DST/SJF/PSA-01/2013-14.

\appendix

\section{Derivation of  $g_i\approx2\sqrt{2}$ in $\D\gg~ \frac{d-2}{2},\ell$ limit, general case.}\label{gi_deri}

Using the large $\D$ limit of the diagonal blocks in \eqref{blockapprox}, we can grind out the positivity criterion analytically at least in the  $\D\gg~ \frac{d-2}{2},\ell$ limit. Step-wise,
\be 
\frac{\partial^{m}_{z}G_{d,\D,\ell}(z)}{m!}{\bigg |}_{z=\frac{1}{2}}=\frac{\sqrt{\pi } \left(2+\frac{3}{\sqrt{2}}\right)^{d/2} \left(12 \sqrt{2}+17\right)^{-\frac{\Delta }{2}} \Delta ^m 2^{-2 d+2 \Delta +\frac{3 m}{2}+3} \Gamma (d+\ell -2)}{\Gamma \left(\frac{d-1}{2}\right) (2)_{m-1} \Gamma \left(\frac{d}{2}+\ell -1\right)}\left[1+O\left(\frac{1}{\D}\right)\right]\,,
\ee
are the block vectors. Now using Leibniz theorem we can calculate  $F_{m,n}=\partial^{n}_{\D}\left(\frac{\partial^{m}_{z}G_{d,\D,\ell}(z)}{m!}\right)$, in a simple closed form expression, 
\be
\begin{split}
F_{m,n}= &\frac{\sqrt{\pi } \left(2+\frac{3}{\sqrt{2}}\right)^{d/2} \left(12-8 \sqrt{2}\right)^{\Delta } 2^{-2 d+\frac{3 m}{2}+3} \Delta ^m (-\Delta )^{-n} \Gamma (d+\ell -2) }{\Gamma \left(\frac{d-1}{2}\right) (2)_{m-1} \Gamma \left(\frac{d}{2}+\ell -1\right)}\\
&U\left(-n,m-n+1,-\Delta  \log \left(12-8 \sqrt{2}\right)\right)\left(1+O(\frac{1}{\D})\right)
\end{split}
\ee
$U$ is HypergeometricU or Tricomi confluent hypergeometric function. Some of the $ \mathbf{K}_{2N+1}(d,\D,\ell)$ are,
\be
\begin{split}
 |\mathbf{K}_{0}(d,\D,\ell)|= &\frac{\sqrt{\pi } 2^{3-2 d} \left(2+\frac{3}{\sqrt{2}}\right)^{d/2} \left(12-8 \sqrt{2}\right)^{\Delta } \Gamma (d+\ell -2)}{\Gamma \left(\frac{d-1}{2}\right) \Gamma \left(\frac{d}{2}+\ell -1\right)}+O(\D^{-1})\,,\\
 |\mathbf{K}_{1}(d,\D,\ell)|= &  \frac{\pi  2^{\frac{15}{2}-4 d} \left(2+\frac{3}{\sqrt{2}}\right)^d \left(12-8 \sqrt{2}\right)^{2 \Delta } \Gamma (d+\ell -2)^2}{\Gamma \left(\frac{d-1}{2}\right)^2 \Gamma \left(\frac{d}{2}+\ell -1\right)^2}+O(\D^{-1})\,,\\ 
 |\mathbf{K}_{2}(d,\D,\ell)|=&  \frac{\pi ^{3/2} 2^{\frac{27}{2}-6 d} \left(2+\frac{3}{\sqrt{2}}\right)^{\frac{3 d}{2}} \left(12-8 \sqrt{2}\right)^{3 \Delta } \Gamma (d+\ell -2)^3}{\Gamma \left(\frac{d-1}{2}\right)^3 \Gamma \left(\frac{d}{2}+\ell -1\right)^3}+O(\D^{-1})\,.
\end{split}
\ee
For the leading order $|\mathbf{K}_i|$, a general formula for large $\D$, then looks like,
\be
|\mathbf{K}_{i}(d,\D,\ell)|= \frac{\pi ^{\frac{i+1}{2}} 2^{\frac{3}{4} (i+1) (i+4)-2 d (i+1)} \left(2+\frac{3}{\sqrt{2}}\right)^{\frac{1}{2} d (i+1)} \left(12-8 \sqrt{2}\right)^{\Delta  (i+1)} \Gamma (d+\ell -2)^{i+1}}{\Gamma \left(\frac{d-1}{2}\right)^{i+1} \Gamma \left(\frac{d}{2}+\ell -1\right)^{i+1}}+O(\D^{-1})\,.
\ee
Now using eq.(\ref{gi}) one can easily compute $g_i$ analytically, to obtain 
\be
g_i \approx 2 \sqrt{2}:~~\forall~i,~~ \D\gg~ \frac{d-2}{2},\ell\,.
\ee 

All the $g_i$'s are positive for large $\D$ for any $d$ and any $\ell$. Therefore we conclude that unitary polytope $\mathbf{U}\{\D_i\}$ is a cyclic polytope in the limit $\D\gg~ \frac{d-2}{2},\ell$

\section{Some examples using known theories}\label{somecheck}
Here we will explicitly check eq.\eqref{stability} for some physical theories  for e.g. 2D and 3D Ising model, taking the first few operators in the spectrum.

\subsection{2d Ising model }

The first few operators in increasing order of $\Delta$ for the 2d Ising model are given by (see 2nd reference in \cite{stuff})
\be
\begin{split}\label{eq:spec2dising}
&\Delta _1=1,~ \Delta _2=2,~ \Delta _3=4,~\Delta _4=4, ~\Delta _5=5,~ \Delta _6=6,~ \Delta _7=6,~ \Delta _8=7,~ \Delta _9=8,~ \Delta _{10}=8\\
&\ell_1=0,~ \ell_2=2,~\ell_3=0,~\ell_4=4,~\ell_5=4,~\ell_6=2,~\ell_7=6,~\ell_8=6,~\ell_9=0,~\ell_{10}=4\,.
\end{split}
\ee

\begin{table}[hbt!]
\centering
\begin{tabular}{|c| | c |}
\hline
$i$ & $\left\langle \vec{G}_{d,\Delta_1,\ell_1},\vec{G}_{d,\Delta_2,\ell_2},~\dots \vec{G}_{d,\Delta_i,\ell_i}\right\rangle$\\ \hline
 1 & $2.9139$ \\ \hline
 2 & $22.6087$ \\ \hline     
 3 & $2162.57$ \\ \hline
 4 & $-15191.9$ \\ \hline
 5 & $1.53684\times 10^6$\\ \hline
 6 & $-5.23824\times 10^7$ \\ \hline
 7 & $-1.85813\times 10^{8}$ \\ \hline
 8 & $-8.39807\times 10^{10}$  \\ \hline
 9 & $ -1.68276\times 10^{12}$ \\ \hline

 \end{tabular}\label{tb:Det2dising}
\quad 
\begin{tabular}{|c| | c |}
\hline
$i$ & $\left\langle \vec{G}_{d,\Delta_1,\ell_1},\vec{G}_{d,\Delta_2,\ell_2},~\dots \vec{G}_{d,\Delta_i,\ell_i}\right\rangle$\\ \hline
 1 & $4.17363$ \\ \hline
 2 & $ 78.5859$ \\ \hline     
 3 & $ 2420.5$ \\ \hline
 4 & $ 249319.$ \\ \hline
 5 & $ 2.51128\times 10^7$\\ \hline
 6 & $ 3.85515 \times 10^9$ \\ \hline
 7 & $ 1.15493 \times 10^{12}$ \\ \hline
 8 & $ -9.71096 \times 10^{12}$  \\ \hline
 9 & $  6.31945 \times 10^{14}$ \\ \hline
 
\end{tabular}
 \label{tb:Det3dising}
\caption{Left: Determinants of first few operators of 2D Ising model using equation (\ref{block}), spectrum (\ref{eq:spec2dising}) and Right: determinants of first few operators of 3D Ising model using \ref{block}, spectrum (\ref{eq:spec3dising}).}
 
\end{table}
Various facets are given below table (\ref{facets_2d_ising}) for the above spectrum for $D=2N=4$  i.e. even dimensional polytopes.  Notice in table (\ref{facets_2d_ising}) that  $\textbf{facet=1,3,6,8,11,12,13,16,17}$ are not part of a cyclic polytopes.

\subsection{3d Ising model }
Similarly for the 3d Ising model \cite{dsd2016}
\be
\begin{split}\label{eq:spec3dising}
&\Delta _1=1.41263,~\Delta _2=3,~\Delta _3=3.82968,~\Delta _4=5.02267,~\Delta _5=5.5091,\\
&\Delta _6=6.42065,~\Delta _7=6.8956,~\Delta _8=7.0758,~\Delta _9=7.2535\\
&\ell_1=0,\ell_2=2,\ell_3=0,\ell_4=4,\ell_5=2,\ell_6=4,\ell_7=0,\ell_8=2,\ell_9=0\,.
\end{split}
\ee

\begin{table}
\be\nonumber
\begin{array}{|c|cccc|}\hline
 \text{facet-1} & \mathbf{G}_{d,\D_0,\ell_0} & \mathbf{G}_{d,\D_1,\ell_1} & \mathbf{G}_{d,\D_2,\ell_2} & \mathbf{G}_{d,\D_4,\ell_4} \\ \hline
 \text{facet-2} & \mathbf{G}_{d,\D_0,\ell_0} & \mathbf{G}_{d,\D_1,\ell_1} & \mathbf{G}_{d,\D_4,\ell_4} & \mathbf{G}_{d,\D_5,\ell_5} \\ \hline
 \text{facet-3} & \mathbf{G}_{d,\D_0,\ell_0} & \mathbf{G}_{d,\D_1,\ell_1} & \mathbf{G}_{d,\D_5,\ell_5} & \mathbf{G}_{d,\D_7,\ell_7} \\ \hline
 \text{facet-4} & \mathbf{G}_{d,\D_0,\ell_0} & \mathbf{G}_{d,\D_1,\ell_1} & \mathbf{G}_{d,\D_7,\ell_7} & \mathbf{G}_{d,\D_8,\ell_8} \\ \hline
 \text{facet-5} & \mathbf{G}_{d,\D_0,\ell_0} & \mathbf{G}_{d,\D_1,\ell_1} & \mathbf{G}_{d,\D_8,\ell_8} & \mathbf{G}_{d,\D_9,\ell_9} \\ \hline
 \text{facet-6} & \mathbf{G}_{d,\D_0,\ell_0} & \mathbf{G}_{d,\D_5,\ell_5} & \mathbf{G}_{d,\D_6,\ell_6} & \mathbf{G}_{d,\D_7,\ell_7} \\ \hline
 \text{facet-7} & \mathbf{G}_{d,\D_1,\ell_1} & \mathbf{G}_{d,\D_2,\ell_2} & \mathbf{G}_{d,\D_4,\ell_4} & \mathbf{G}_{d,\D_5,\ell_5} \\ \hline
 \text{facet-8} & \mathbf{G}_{d,\D_1,\ell_1} & \mathbf{G}_{d,\D_2,\ell_2} & \mathbf{G}_{d,\D_5,\ell_5} & \mathbf{G}_{d,\D_7,\ell_7} \\ \hline
 \text{facet-9} & \mathbf{G}_{d,\D_1,\ell_1} & \mathbf{G}_{d,\D_2,\ell_2} & \mathbf{G}_{d,\D_7,\ell_7} & \mathbf{G}_{d,\D_8,\ell_8} \\ \hline
 \text{facet-10} & \mathbf{G}_{d,\D_1,\ell_1} & \mathbf{G}_{d,\D_2,\ell_2} & \mathbf{G}_{d,\D_8,\ell_8} & \mathbf{G}_{d,\D_9,\ell_9} \\ \hline
 \text{facet-11} & \mathbf{G}_{d,\D_2,\ell_2} & \mathbf{G}_{d,\D_4,\ell_4} & \mathbf{G}_{d,\D_5,\ell_5} & \mathbf{G}_{d,\D_7,\ell_7} \\ \hline
 \text{facet-12} & \mathbf{G}_{d,\D_2,\ell_2} & \mathbf{G}_{d,\D_4,\ell_4} & \mathbf{G}_{d,\D_7,\ell_7} & \mathbf{G}_{d,\D_8,\ell_8} \\ \hline
 \text{facet-13} & \mathbf{G}_{d,\D_2,\ell_2} & \mathbf{G}_{d,\D_4,\ell_4} & \mathbf{G}_{d,\D_8,\ell_8} & \mathbf{G}_{d,\D_9,\ell_9} \\ \hline
 \text{facet-14} & \mathbf{G}_{d,\D_4,\ell_4} & \mathbf{G}_{d,\D_5,\ell_5} & \mathbf{G}_{d,\D_7,\ell_7} & \mathbf{G}_{d,\D_8,\ell_8} \\ \hline
 \text{facet-15} & \mathbf{G}_{d,\D_4,\ell_4} & \mathbf{G}_{d,\D_5,\ell_5} & \mathbf{G}_{d,\D_8,\ell_8} & \mathbf{G}_{d,\D_9,\ell_9} \\ \hline
 \text{facet-16} & \mathbf{G}_{d,\D_5,\ell_5} & \mathbf{G}_{d,\D_6,\ell_6} & \mathbf{G}_{d,\D_7,\ell_7} & \mathbf{G}_{d,\D_9,\ell_9} \\ \hline
 \text{facet-17} & \mathbf{G}_{d,\D_5,\ell_5} & \mathbf{G}_{d,\D_7,\ell_7} & \mathbf{G}_{d,\D_8,\ell_8} & \mathbf{G}_{d,\D_9,\ell_9} \\ \hline
\end{array}
\ee 
\caption{Various facets for $D=2N=4$ for 2d ising model, where $\mathbf{G}_{d,\D_i,\ell_i}$ are the block vectors for $D=2N=4$ and $\mathbf{G}_{d,\D_0,\ell_0}=(1,0,0,0,0)$}
\label{facets_2d_ising}
\end{table}

\subsection{Different ordering of combinations of $\D$ and $\ell$}
In above discussions we have ordered the spectrum in increasing value of $\D$. It may be possible that ordering in terms of say twist $\Delta-\ell$, will lead to a cyclic polytope even for $O(1)$ values of $\D$. In this appendix we investigate one such case namely ordering in terms of increasing value of $\D-\ell$, see table (\ref{tb:Det2disingDmpl}). For the 2d Ising model, there are degenerate values of $\D-\ell$. From the spectrum eq ( \ref{eq:spec2dising}) (we consider the first 9 operators) there are $3!\times 3! \times 3!=216$ degeneracy. Out of all $216$ orderings there are only $4$ combinations that give positive determinants and these ordering are 

\be\label{sp4}
\begin{split}
1)~&\D_1=2,\D_2=4,\D_3=6,\D_4=7,\D_5=1,\D_6=5.,\D_7=8,\D_8=6.,\D_9=4.\\
~~&\ell_1=2,\ell_2=4,\ell_3=6,\ell_4=6,\ell_5=0,\ell_6=4,\ell_7=4,\ell_8=2,\ell_9=0\\
2)~&\D_1=2,\D_2=4,\D_3=6,\D_4=7,\D_5=1,\D_6=5,\D_7=6.,\D_8=4.,\D_9=8.\\
~~&\ell_1=2,\ell_2=4,\ell_3=6,\ell_4=6,\ell_5=0,\ell_6=4,\ell_7=2,\ell_8=0,\ell_9=4\\
3)~&\D_1=4,\D_2=6,\D_3=2,\D_4=7,\D_5=1,\D_6=5.,\D_7=8,\D_8=6,\D_9=4.\\
~~&\ell_1=4,\ell_2=6,\ell_3=2,\ell_4=6,\ell_5=0,\ell_6=4,\ell_7=4,\ell_8=2,\ell_9=0\\
4)~&\D_1=4,\D_2=6.,\D_3=2,\D_4=7,\D_5=1,\D_6=5.,\D_7=6,\D_8=4.,\D_9=8.\\
~~&\ell_1=4,\ell_2=6,\ell_3=2,\ell_4=6,\ell_5=0,\ell_6=4,\ell_7=2,\ell_8=0,\ell_9=4
\end{split}
\ee
Even though these ordering give positive determinants but we did not find any general pattern of the ordering. Also in the table, we have ordered degenerate operators with respect to increasing $\D$. Other possible orderings do not change the conclusion.  It may be possible that some complicated combination of $\D$ and $\ell$ can lead to cyclic polytope structure in even small $\D$ range which we have not found so far.

\begin{table}[hbt!]
\centering
\begin{tabular}{|c| | c |}
\hline
$i$ & $\left\langle \vec{G}_{d,\Delta_1,\ell_1},\vec{G}_{d,\Delta_2,\ell_2},~\dots \vec{G}_{d,\Delta_i,\ell_i}\right\rangle$\\ \hline
 1 & $5.72525$ \\ \hline
 2 & $183.694$ \\ \hline     
 3 & $33303.3$ \\ \hline
 4 & $-1.32982\times 10^6$ \\ \hline
 5 & $1.08811\times 10^8$\\ \hline
 6 & $1.28685\times10^{11}$ \\ \hline
 7 & $6.96021\times 10^{10}$ \\ \hline
 8 & $ -8.39807\times10^{10}$  \\ \hline
 9 & $ -1.80068\times10^{12}$ \\ \hline

 \end{tabular} 
\quad 
\begin{tabular}{|c| | c |}
\hline
$i$ & $\left\langle \vec{G}_{d,\Delta_1,\ell_1},\vec{G}_{d,\Delta_2,\ell_2},~\dots \vec{G}_{d,\Delta_i,\ell_i}\right\rangle$\\ \hline
 1 & $5.72525$ \\ \hline
 2 & $183.694$ \\ \hline     
 3 & $33303.3$ \\ \hline
 4 & $9.91382\times 10^6$ \\ \hline
 5 & $1.37549\times 10^9$\\ \hline
 6 & $1.28685\times 10^{11}$ \\ \hline
 7 & $2.97071\times 10^{14}$ \\ \hline
 8 & $1.36497\times 10^{15}$  \\ \hline
 9 & $ 1.80068\times 10^{12}$ \\ \hline

\end{tabular}
 
\caption{Left: Determinants of first few operators of 2D Ising model using equation (\ref{block}), spectrum (\ref{eq:spec2dising}) in increasing order of $\D-\ell$, here we have ordered the degenerate ordering in increasing ordering of $\D$  and Right: determinants of first few operators of 2D Ising model using \ref{block}, spectrum (\ref{eq:spec2dising}) in increasing order of $\D-\ell$ using spectrum (1) of eq (\ref{sp4}).} 
 \label{tb:Det2disingDmpl}
\end{table}

\subsubsection*{Remarks}\label{somecheck_remark}

 The numbers listed in table 5 are found using (\ref{block}). Some determinants are negative which show that $P(c_\ell)$ is not a cyclic polytope. However, we get positive numbers if we use (\ref{blockapprox}).  Also if one calculates  for each $\ell$ separately then all the determinants are positive. This means that each spin-$\ell$ vectors lead to cyclic polytopes while the weighted Minkowski sum is not. However, as we have shown earlier, the approximate block in eq.(\ref{blockapprox}) is in very good agreement numerically with the exact block (\ref{block}). Therefore, in some approximate sense $P(c_\ell)$ is also a cyclic polytope. \\
 Also we have explicitly calculated the faces of the polytopes e.g. table (\ref{facets_2d_ising}) and we find that $P(c_\ell)$ is not a cyclic polytopes and  for each $\ell$ separately they from cyclic polytopes. And we find that $P(c_\ell)$ is cyclic polytopes if we use (\ref{blockapprox}).

\section{Analytic kink analysis}\label{kink}
Here will briefly elaborate on our attempts to track down the location of the kink, in the cyclic polytope picture in section 7, analytically. By plugging in the asymptotic blocks to the $1/\D_S$ order which can be easily found using eq. (2.3), one can easily find that the the largest zero of $det_3$ (obtained by expanding the determinant to $O(\D_S^2)$) which we will denote by $\zeta_0^{(3)}$ is given by
\be
\zeta_0^{(3)}=3\sqrt{2}\Dphi+3-\frac{3}{\sqrt{2}}\,.
\ee
On the other hand, the location of the largest zero of $det_4$ is given by a 5-th order polynomial equation (obtained by expanding the determinant to $O(\D_S^2)$) whose roots cannot be found analytically. However, we are interested in crossing of the zeros of $det_3$ and $det_4$ so we will just put the largest zero of $det_4$ as 
\be
\zeta_0^{(4)}=\zeta_0^{(3)}+\delta\,,
\ee
and Taylor expand in $\delta$ to obtain a simpler equation. This equation now (in the context of the analysis in section 7) depends on the $\D_8$. We can expand in the limit where $\D_8\gg 1$ and find an approximate equation whose (positive) root gives us the value of $\Dphi$ at the kink. This works out to be 
\be
\Dphi=\frac{-33-24\sqrt{2}+\sqrt{4277+3024\sqrt{2}}}{16(17+12\sqrt{2})}\approx 0.0469974\,,
\ee
using which we find
\be
\D_S\approx 1.07808\,.
\ee
Considering the fact that we have used only the $1/\D_S$ approximation of the block and we are trying to track down a kink at $\D_S\sim O(1)$, it is not surprising that we do not get a very precise location which is closer to $(\Dphi,\D_S)=(0.08,1.2)$ but nonetheless this would be a good starting point to see if a systematic solution can be worked out by retaining higher order $1/\D_S$ terms.

\section{Adding spin and kink}\label{ap:kink_spin}
\begin{figure}[hbt!]
  \centering
  \begin{subfigure}[b]{0.45\linewidth}
    \includegraphics[width=\linewidth]{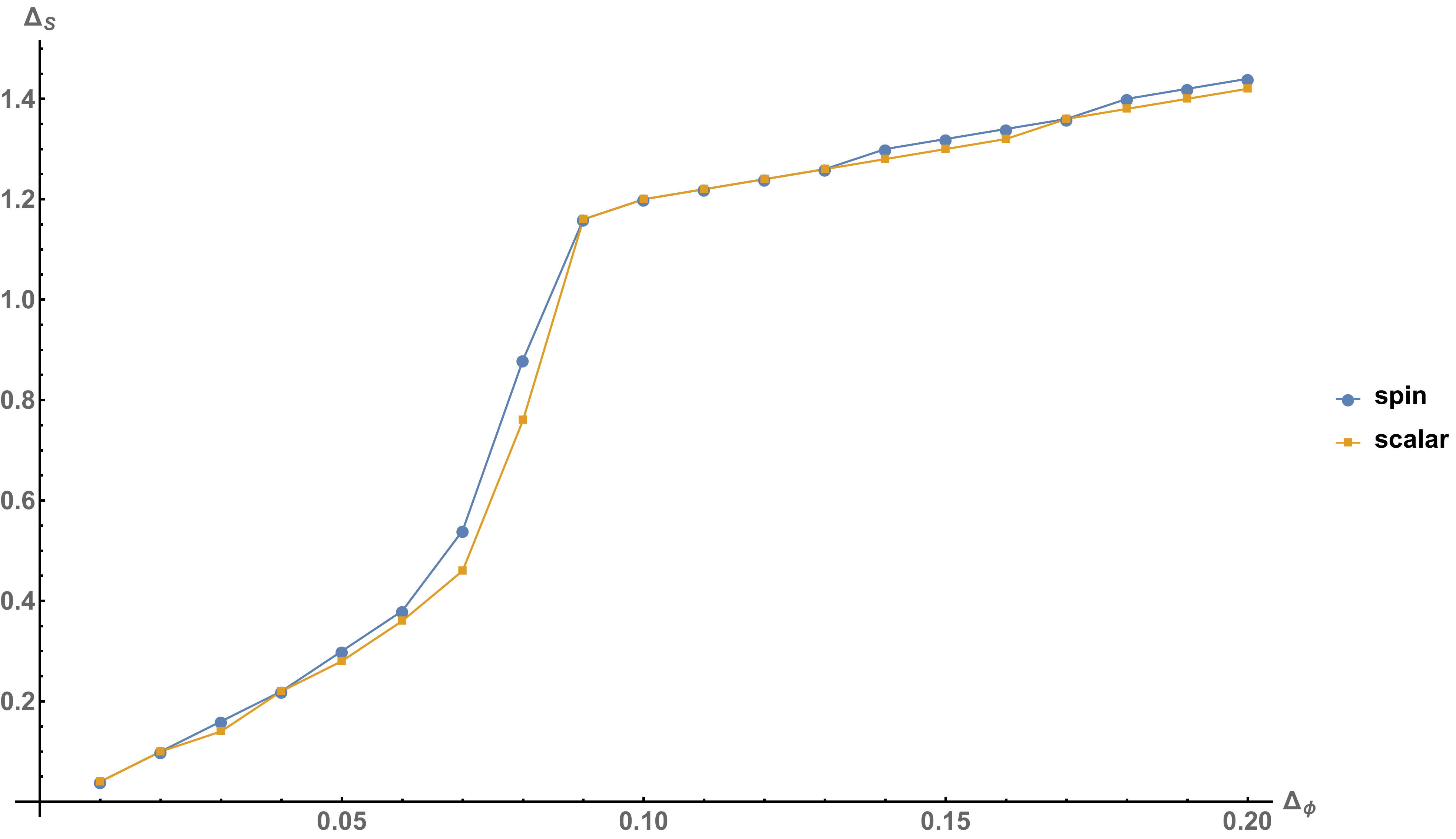}
    \caption{$\D_S$ vs $\Dphi$ for $d=2$ using linear programming, 10 derivative in eq (\ref{cross1}).}
  \end{subfigure}
  \begin{subfigure}[b]{0.45\linewidth}
    \includegraphics[width=\linewidth]{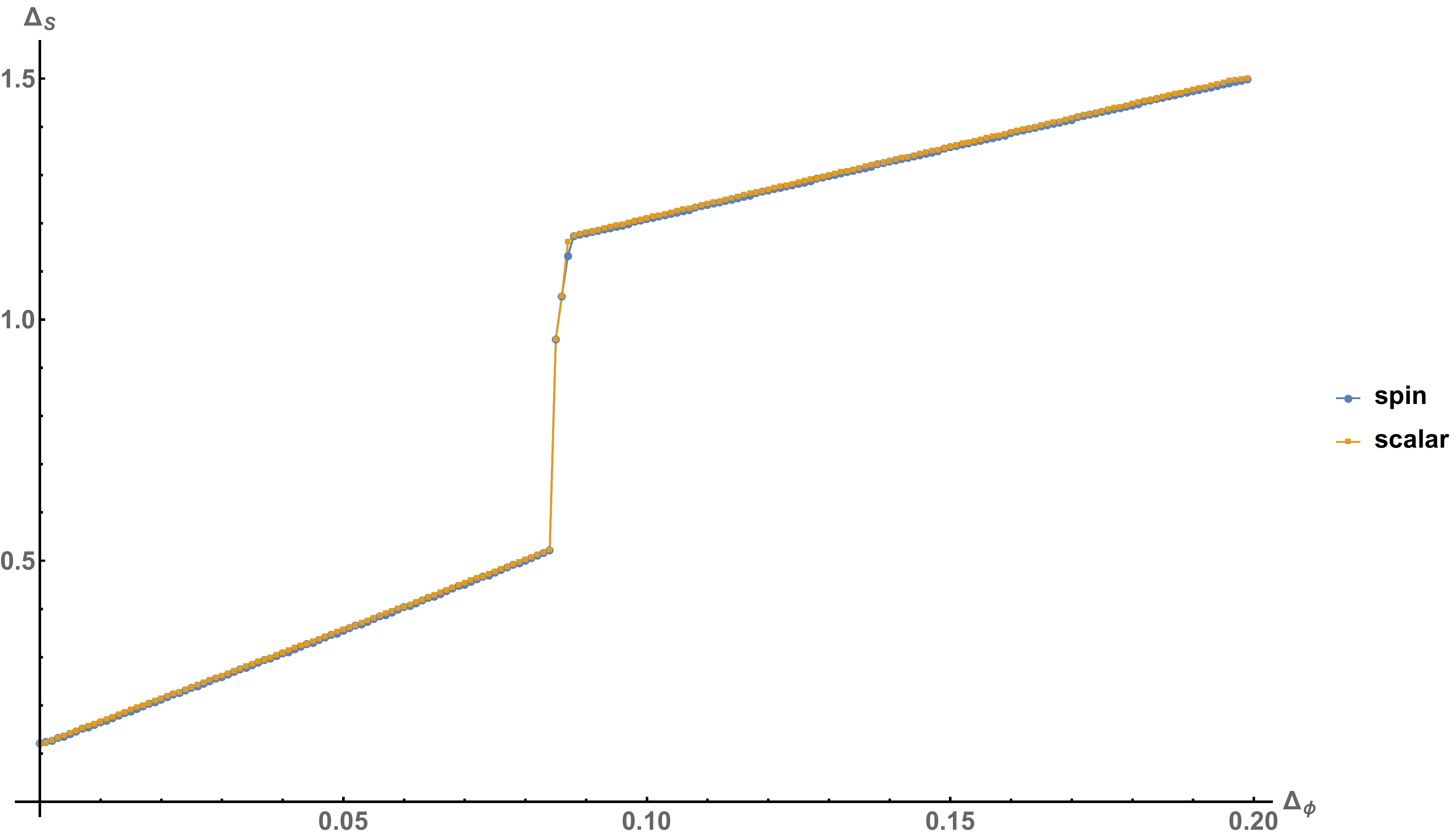}
    \caption{$\D_S$ vs $\Dphi$ for $d=2$ using $N=2$ analysis.}
   \end{subfigure}
  \caption{Bound for the leading scalar dimension including spins in the spectrum. In each plot, the blue line is  obtained considering up to spin $\ell=6$ in the analysis. Yellow line is only considering the scalar in the spectrum }
  \label{fig:llp_spin}
\end{figure}

In this appendix we consider the effect of adding spin to the analysis of bounds on the leading scalar dimension $\D_S$. In the diagonal limit, it seems that position of spin does not move much, even if we add spins. See figure (\ref{fig:llp_spin}). We will study this effect in detail in an upcoming work \cite{YSWZ}.

\section{Issues with scalars}\label{smallD}
It is possible that for small $\D$ is it possible to get the determinants  $\left\langle \vec{G}_{d,\Delta_1,\ell_1},\vec{G}_{d,\Delta_2,\ell_2},~\dots \vec{G}_{d,\Delta_D,\ell_D}\right\rangle<0$. For the 2d and 3d Ising models, this does not happen for each spin separately. We rigorously checked the determinants scanning various $\D$ in Mathematica. For example, for only scalars \footnote{Only scalars create problems near the unitary value of $\D$ as is clear from figure (\ref{fig:scalarblock},  \ref{fig:spin2block})} if we assume spectrum like $\D_j=\frac{d-2}{2}+a\times j$ then to have positivity the minimum \footnote{One can fine grain it by increasing precision in Mathematica which we have not done here.} value of $a$ is
\be
\begin{split}
 &a>0.0326,~d=2\\ 
 &a>0.1906,~d=3 \\ 
 &a>0.2395,~d=4\,,
\end{split}
\ee
for $D=2N+1=7$ dimensional polytope.

Note that from definition of $F_{mn}$ eq (\ref{Fmn})
\be
\frac{\partial^{m}G_{d,\D_j,\ell}(z)|_{z=1/2}}{m!}=\sum_{n}(\D_j-\D)^{n} \frac{F_{m,n}}{n!}
\ee
We want to analyse the behaviour of $\left\langle \vec{G}_{d,\Delta_1,\ell},\vec{G}_{d,\Delta_2,\ell},~\dots \vec{G}_{d,\Delta_D,\ell}\right\rangle$ assuming the spectrum for fixed spin $\ell$, $\D_j=\D+\d_j$, where $\d_j$ is small and $\d_{j+1}>\d_{j}$ \textit{i.e} this spectrum respects the increasing ordering in $\D_j$
\be
\left\langle\vec{G}_{d,\Delta_{1},\ell}, \vec{G}_{d,\Delta_{2},\ell}, \cdots, \vec{G}_{d,\Delta_{D},\ell}\right\rangle=\a_{D}(d,\D,\ell)\prod_{i<j}\left(\d_{j}-\d_{i}\right)+\mathcal{O}\left(\d^{\frac{D(D-1)}{2}+1}\right)~;~D>2
\ee
where 
\be
\a_{D}(d,\D,\ell)= det \left( \begin{array}{ccccc}{F_{0,0}} & {F_{1,0}} & { . .} & { . .} & {F_{2 N+1,0}} \\ {F_{0,1}} & {F_{1,1}} & { . .} & {. . } & { . .} \\ { . .} & { . .} & {\frac{F_{i, j}}{j!}} & { . .} & { . .} \\ { . .} & { . .} & { . .} & { . .} & { . .} \\ {\frac{F_{0,D}}{D!}} & { . .} & { . .} & { . .} & {\frac{F_{D,D}}{D!} }\end{array}\right)\,.
\ee
So we conclude that if $\a_{D}(d,\D,\ell)$'s are positive whatever small number $\d_j$ is as long as they are ordered (\textit{i.e} $\d_{j+1}>\d_{j}$) the cyclicity holds. Also note that $\a_D$ defined in \cite[eq (5.15)]{nima} is equal to our $\a_D(d,\D,\ell)$
\be
\a_{D}(d=1,\D=0,\ell=0)=\a_D
\ee
We find \be
|\mathbf{K}_D(d,\D,\ell)|=\left(\prod_{j=0}^{D}j!\right)\a_{D}(d,\D,\ell)\,.
\ee
Knowing $g_1>0, g_2>0$ it is then recursively true that for positivity of $D$ dimensional polytope it is sufficient that $g_i>0,~ \forall i,~1\leq i \leq D$. 

\subsection*{Note}
This analysis shows how $\mathbf{K}_i(d,\D,\ell)$ appears. We have checked analytically that our definition of $g_i$'s in terms of $\mathbf{K}_i(d,\D,\ell)$  eq (\ref{gi}) is same given in \cite{nima}. Even though we don't have first hand proof of it. We guess the formula by observing first three $g_i$'s and verified it till $g_8$ that it is true.

 Also it is important to mention that Mathematica has some numerical issues evaluating the determinants eq (\ref{stability}). It gives very small negative numbers ($\sim 10^{-50}$) or very big negative numbers ($\sim 10^{50}$) when the spacing $\d_j$ is very small even if in the region of $\D$ for which $g_i$'s are positive. That this should not happen is clear from the above analysis.

\end{document}